\documentclass[final]{ias2}

\usepackage{graphicx,amsmath,amssymb} 
\usepackage{multirow}
\usepackage{array} 

\usepackage{hyperref} 

\newcommand \ra  {\rightarrow}

\newcommand \w  {\omega}

\newcommand \A {\alpha}

\newcommand \lc {\langle}
\newcommand \rc {\rangle}
\newcommand \prt {\partial}

\newcommand \bvec{\left( \begin{array}{c} }
\newcommand \evec{\end{array} \right)}
\newcommand \tr {\mbox{{ Tr}}}

\newcommand \bea{\begin{eqnarray} }
\newcommand \eea{\end{eqnarray} } 
\newcommand \nn {\nonumber}
\newcommand {\be} {\begin{equation}}
\newcommand {\ee} {\end{equation}}

\newcommand {\mbx} {\mbox{}}

\newcommand {\ata} {& \times &}

\begin{document}

\markboth{Jet modification in the next decade}{Abhijit Majumder}

\title{Jet modification in the next decade: a pedestrian outlook}

\author[sin]{Abhijit Majumder} 
\email{abhijit.majumder@wayne.edu}
\address[sin]{Department of Physics and Astronomy, Wayne State University, Detroit, Michigan 48201, USA}

\begin{abstract}
In this review, intended for non-specialists and beginners, we recount the current status of the theory of jet modification in dense matter. We commence with an outline of the ``traditional'' observables which may be calculated without recourse to event generators. These include single and double hadron suppression, nuclear modification factor versus reaction plane etc. 
All of these measurements are used to justify both the required underlying physical picture of jet modification 
as well as the final obtained values of jet transport coefficients. 
This is followed by a review of the more modern observables which have arisen with the ability to reconstruct full jets, and the challenges faced therein. 
This is followed by a preview of upcoming theoretical developments in the field and an outlook on how the interface between these developments, 
phenomenological improvements, and upcoming data will allow us to 
quantitatively determine properties of the medium which effect the modification of hard jets. 
\end{abstract}


 
\maketitle

\section{Introduction and history} ~\label{intro}

At the time of this writing, there have been three heavy-ion runs at the Large Hadron Collider (LHC) at $\sqrt{s} = 2.75$TeV per nucleon pair, at the same time there have been twelve runs at the Relativistic Heavy-Ion Collider (RHIC), mostly at the highest energy of $\sqrt{s} = 200$GeV per nucleon pair. 
The simultaneous running of these facilities at distinct energy scales has led to both dramatic improvements in our 
theoretical understanding of the modification of hard jets in the quark gluon plasma (QGP), 
as well as to the origin of novel experimental techniques and observables that may be applied to jet modification at 
both these energy scales. Experimental results from several years of running at RHIC, 
established the presence of three regions in the spectrum of produced 
particles~\cite{Arsene20051,Back200528,Adams2005102,Adcox2005184}. The soft bulk, which includes particles produced with transverse momenta 
$p_{T}  \lesssim  2$GeV, the hard sector with $p_{T} \gtrsim 8$ GeV, and the intermediate $p_{T}$ region with $2$GeV $\lesssim p_{T} \lesssim 8$GeV 
(all the equalities should be understood to be approximate).

Within a few years of RHIC running, the theoretical understanding of the dynamics of the soft sector, 
via the model of viscous hydrodynamics began to reach a consensus~\cite{Kolb:2003dz,Teaney:2001av}. 
Not withstanding issues related to the rapid thermalization of the produced matter, or the role of fluctuations, the soft sector at the LHC has so far remained a straightforward extension of this scheme. The fate of the hard sector has been rather different; prior to the start of the LHC, there remained multiple approaches, which included both weak coupling approaches based on perturbative QCD (pQCD)~\cite{Gyulassy:2001nm,Gyulassy:2000er,Gyulassy:2000fs,Gyulassy:1999zd,Armesto:2003jh,Salgado:2003gb,Wiedemann:2000tf,Wiedemann:2000za,Arnold:2008vd,Arnold:2002ja,Arnold:2001ba,Arnold:2001ms,Wang:2001ifa,Guo:2000nz,Majumder:2007hx,Majumder:2007ne,Majumder:2009ge,Idilbi:2008vm}, as well as strong coupling approaches based on anti-de-Sitter space/Conformal Field Theory (AdS/CFT) conjecture~\cite{CasalderreySolana:2007qw,Gubser:2009sn,Gubser:2006bz,Herzog:2006gh}. 
For extended reviews on the comparisons between these approaches see Refs.~\cite{CasalderreySolana:2007zz,Majumder:2010qh,Armesto:2011ht}.
For a more concise and focussed review on the comparisons between the different pQCD based approaches see Ref.~\cite{Majumder:2007iu}. 
\begin{figure}[ht]
\begin{center}
\includegraphics[width=0.9\columnwidth]{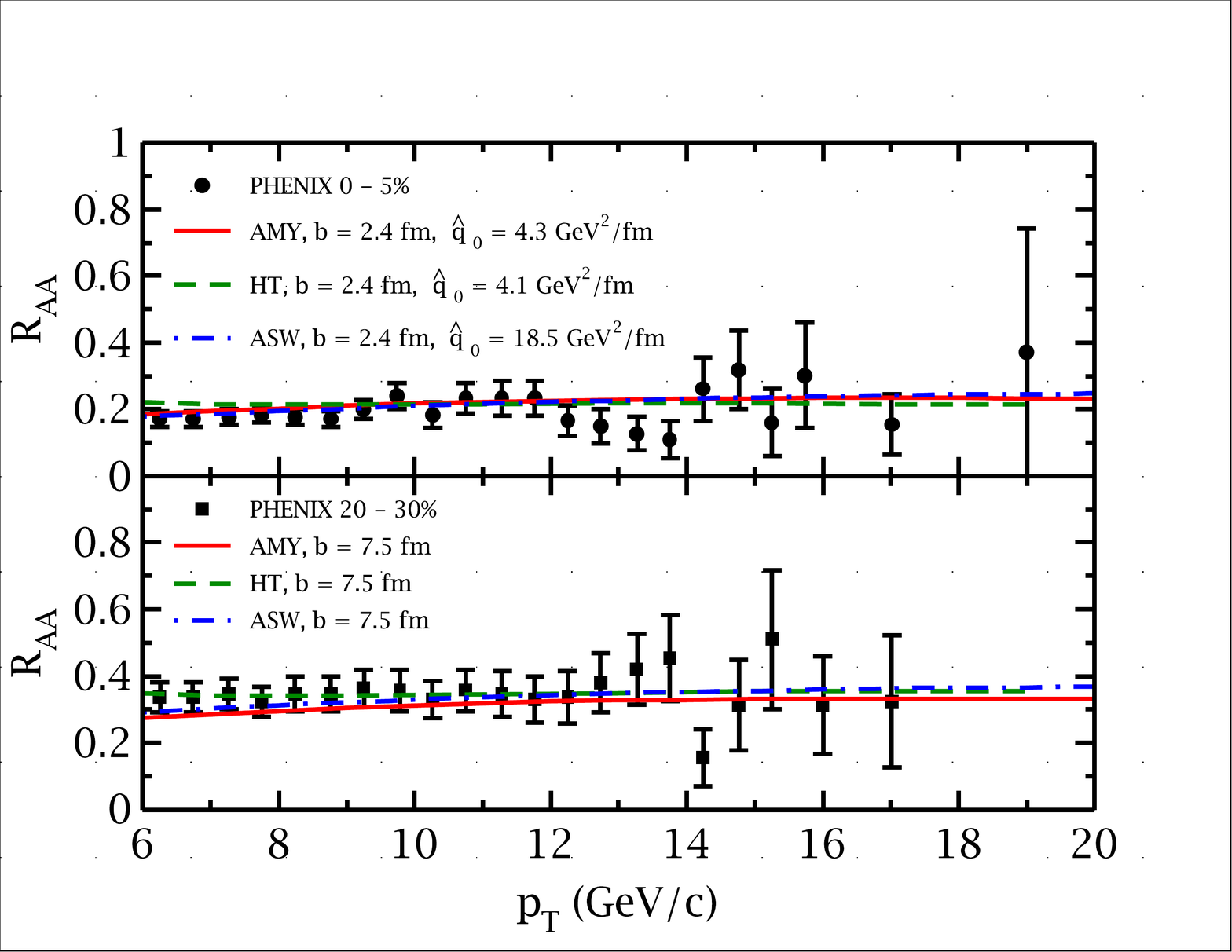}
\caption{The nuclear modification factor $R_{AA}$ as a function of the transverse momentum $p_{T}$  of the detected hadron, at two centralities: 
(0-5\%) most central and (20-30\%) semi-peripheral collisions, at RHIC energies of $\sqrt{s} = 200$GeV per nucleon pair. Figure reproduced with permission 
from Ref.~\cite{Bass:2008rv}, legends slightly modified. }
\label{DJMON-plot}
\end{center}
\end{figure} 

Within the various perturbative approaches, attempts at a serious comparison between three different schemes and simultaneously with data, led to the work of Ref.~\cite{Bass:2008rv}, where the Armesto-Salgado-Wiedemann (ASW), the Arnold-Moore-Yaffe (AMY), and a variant of the Higher-Twist (HT) schemes were constrained to compute leading particle suppression on an identical fluid-dynamical background (that of Ref.~\cite{Nonaka:2006yn}). In all approaches the jet quenching parameter $\hat{q}$, defined as  the mean transverse momentum squared per unit length, acquired by a hard parton, depended on an intrinsic property of the medium such as 
the local temperature $T$, entropy density $s$, or the energy density $\epsilon$ (one should point out that in the AMY scheme, the actual input parameter is the in-medium strong coupling 
constant $\A_{s}$, and $\hat{q}_{0}$ is calculated as a byproduct). Figure~\ref{DJMON-plot}  contains plots of the nuclear modification factor $R_{AA}$ defined as, 
\bea
R_{AA} (b_{min}, b_{max}) &=& \frac{ \frac{d^{2} N_{AA} (b_{min},b_{max}) }{ d^2 p_{T} dy }  }{   \langle N_{bin} (b_{min}, b_{max}) \rangle \frac{d^{2} N_{pp} }{d^{2} p_{T} dy}  }
\eea
In the equation above, $\frac{d^{2} N_{AA} (b_{min},b_{max}) }{ d^2 p_{T} dy } $ represents the yield of particles in a nucleus-nucleus collision, in bins of transverse momentum $p_{T}$ and rapidity $y$, in 
a range of impact parameters that qualify a certain centrality.  The factor $d^{2} N_{pp}/d^{2}p_{T} dy$, in the denominator, 
represents the yield in $p$-$p$ collisions in the same $p_{T}$ and $y$ bins, and $N_{bin}$ represents the number of binary nucleon nucleon collisions in the same range of 
impact parameter as that included in the numerator.
While all the approaches, reproduced the centrality and $p_{T}$ dependence as seen in the RHIC data, the underlying input parameter $\hat{q}_{0}$, which 
refers to the maximum value of $\hat{q}$, achieved at the center of the most central collision at the start of the hydro simulation ($\tau = 0.6$fm/c), turned out to be different 
by a factor of 5, between the ASW and the other approaches, see Fig.~\ref{DJMON-plot}.

This scenario changed quickly with the start up of the LHC heavy-ion program. The same calculations were then extrapolated parameter-free to LHC energies (once the functional dependence of $\hat{q}$ on $T$ or $s$ has been set at RHIC energy, there are no more parameters to tune in comparisons with LHC data). 
We should point out that the qualification of ``parameter-free'' only refers to the jet sector, 
additional parameters do enter in the modeling of the bulk matter, and in the extrapolation from RHIC to LHC energies. 
We present here the results of such a comparison from Ref.~\cite{Muller:2012zq}. 
\begin{figure}[!ht]
\begin{flushleft}
$\mbx$\hspace{-0.5in}\includegraphics[width=0.7\columnwidth]{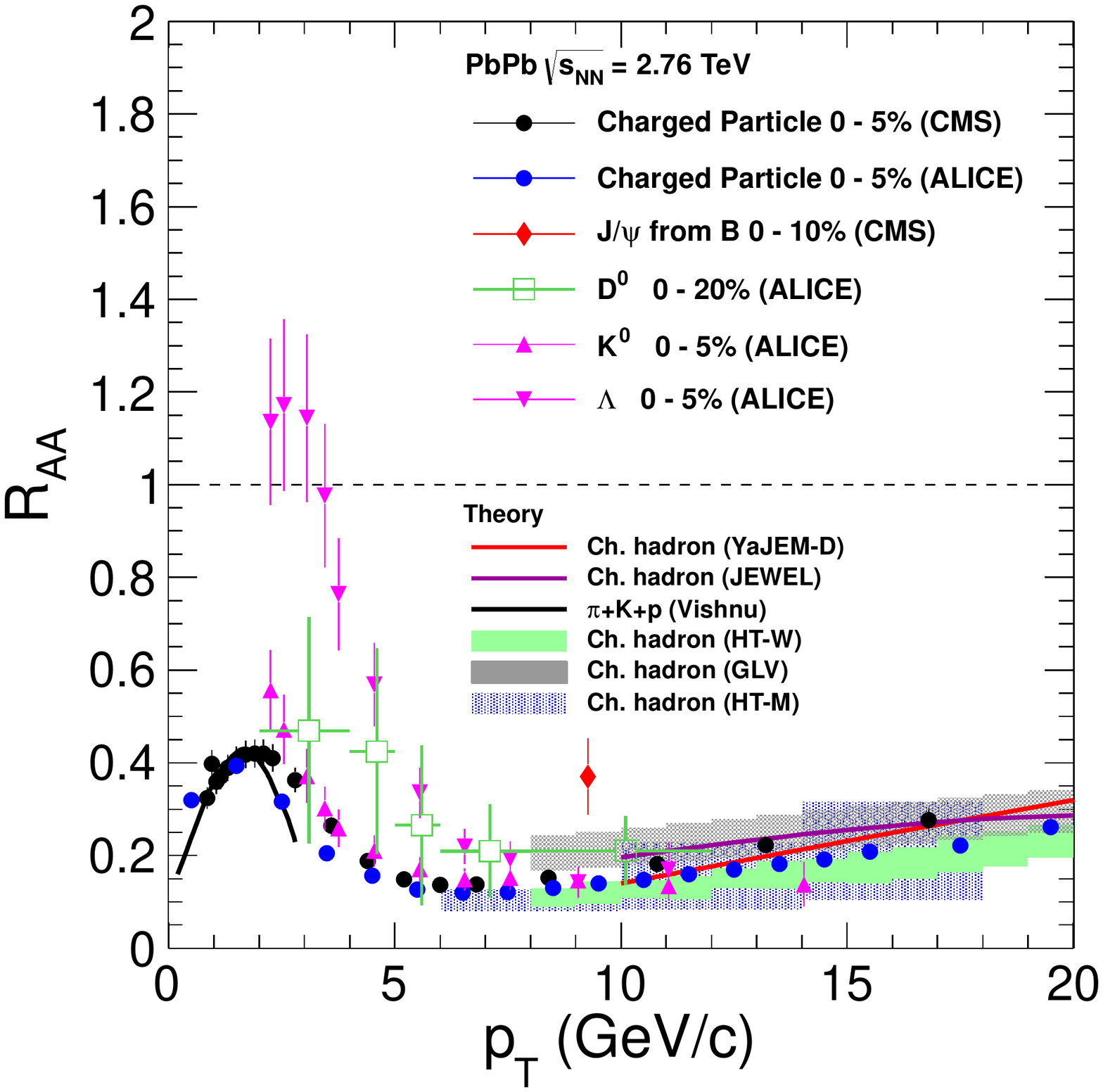}~\hspace{-1in}\includegraphics[width=0.7\columnwidth]{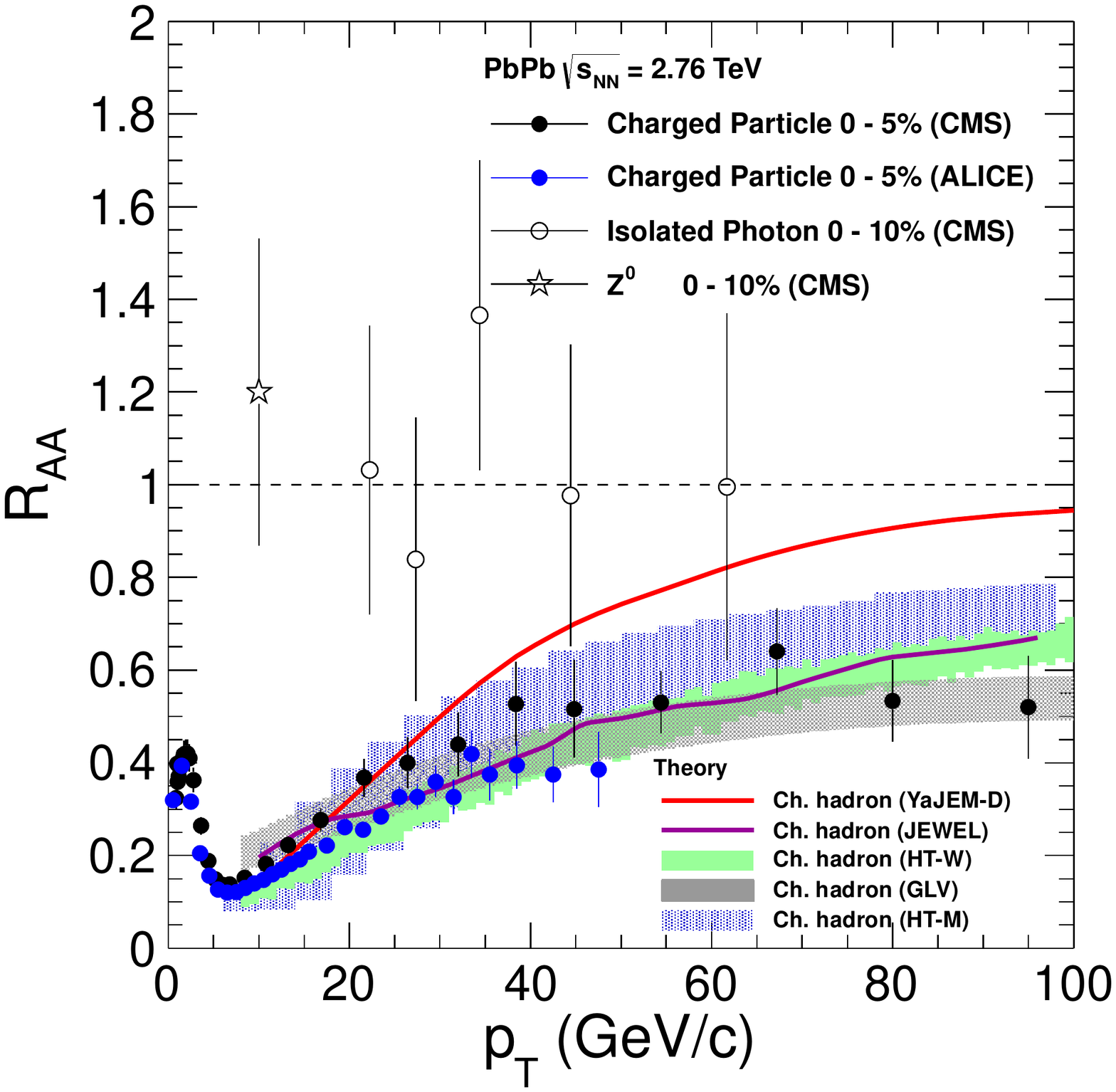}
\caption{The nuclear modification factor $R_{AA}$ as a function of the transverse momentum $p_{T}$  of the detected hadron, for the most central 
(0-5\%) collisions, at LHC energies of $\sqrt{s} = 2.75$TeV per nucleon pair. Figure obtained via private communication from the authors of Ref.~\cite{Muller:2012zq}, legends slightly modified. Experimental data are from the CMS and ALICE collaborations, reproduced here with permission from CERN.}
\label{MSW-plot}
\end{flushleft}
\end{figure} 
In the Fig.~\ref{MSW-plot}, the nuclear modification factor $R_{AA}$ has been plotted against the transverse momentum $p_{T}$ of the detected hadron. Lower momenta are 
contained in the plot on the left hand side, while higher momentum are in the plot on the right hand side. The lines marked JEWEL~\cite{Zapp:2008af,Zapp:2008gi,Zapp:2011ya} and YAJEM-D~\cite{Renk:2010qx,Renk:2010zx} represent calculations from 
event generators, and will be discussed later. The lines marked HT-M, HT-W and GLV represent non-event-generator calculations which have been extrapolated from RHIC data. 
In this particular version, the GLV calculations are not performed on a fluid dynamical medium (this along with other effects such as running coupling have been introduced recently, 
leading to a better agreement between GLV based calculations and LHC data). The AMY calculations do not appear on this plot and 
have since been redone with additional effects such as the inclusion of vacuum medium interferences 
in order to explain the features seen in the LHC data. Calculations based on strong coupling approaches, as well as pQCD based 
calculations that required a top value of $\hat{q}_{0} > 5$GeV$^{2}$/fm at RHIC energies have been disfavored by the LHC data.
Both HT based calculations were done on fluid dynamical simulations tuned to describe RHIC and LHC data. The hydro calculations on which the HT-M calculations are carried out are 
represented by the line denoted as $\pi+K+P$ (Vishnu) on the left plot.

Based on the ambit of successful pQCD based models, one may outline the basic features that any realistic pQCD based calculation of energy loss must possess. These will 
be outlined in the subsequent section (Sec. II). This will be followed by a discussion of how these compare with the full range of single and two-particle observables. With the energy reach of the LHC one may carry out measurements of full jets as opposed to leading particles. This has led to an entirely new set of observables, the theoretical issues associated with these will be described in Sec. III. In Sec. IV, we describe the new theoretical developments that are beginning to appear, and those that will appear in the coming years. Concluding discussions are presented in Sec. V.

\section{One and Two Particle Observables}\label{1-particle}

Within the framework afforded by the factorization theorems of perturbative QCD, high-$p_{T}$ single and double particle inclusive cross sections represent the most well defined and controlled calculations. In $p$-$p$ collisions, the inclusive cross section of a single inclusive high $p_{T}$ pion may be expressed in factorized form as, 
\bea
\frac{d^{2} \sigma}{ d p_{T}^{2} dy } &=& \int dx_{a} dx_{b} G(x_{a},\mu^{2}) G(x_{b},\mu^{2}) \frac{d \hat{\sigma}}{ d \hat{t} } \frac{D^{\pi}(z, \mu^{2})}{ \pi z} ,
\eea 
where, $G(x_{a},\mu^{2}) [G(x_{b}, \mu^{2})]$ represent parton distribution functions (PDFs) of the incoming protons with forward momentum fractions $x_{a},x_{b}$, these are factorized from the hard partonic cross section $d \hat{\sigma}/{d \hat{t}}$ and the final fragmentation function $D(z,\mu^{2})$ where the pion takes away a fraction $z$ of the outgoing parton's energy. The scale 
$\mu^{2}$ is the factorization scale, and at this order, can be chosen to be the hard scale of the process, denoted as $Q^{2}$. The above formula, receives corrections which are power suppressed by the hard scale of the process: the first correction is of the order $\Lambda^{2} / Q^{2}$, where $\Lambda^{2}$ is a scale associated with soft higher-twist matrix elements within the proton (or the bulk portion of the final state of a $p$-$p$ collision). The hard scale of the process $Q^{2}$ is of the order of the measured 
$p_{T}^{2}$ of the produced pion. Thus at high $p_{T}$, such corrections are negligible.

The validity of the above factorized form for $p$-$p$ collisions has been established via factorization theorems~\cite{Collins:1983ju,Collins:1985ue,Collins:1988ig,Collins:1989gx}.
For the case of high-$p_{T}$ hadron production in a heavy-ion collision, no such theorem has so far been demonstrated; however, there is wide consensus that a similar theorem may exist and 
one may decompose the cross section as, 
\bea
\frac{d^{2} \sigma_{AA}}{ d p_{T}^{2} dy } &=& \int_{b_{min}}^{b_{max}} d^{2} b d^{2} r_{\perp} t_{AB}  (\vec{r}_{\perp},\vec{b})  
\int dx_{a} dx_{b} S_{A}(x_{a}, \vec{r}_{\perp},\vec{b}_{\perp}, \mu^{2})  \label{HIC-cross-section}\\ 
\ata S_{B}(x_{a}, \vec{r}_{\perp},\vec{b}_{\perp}, \mu^{2}) 
G(x_{a},\mu^{2}) G(x_{b},\mu^{2}) \frac{d \hat{\sigma}}{ d \hat{t} } \frac{\widetilde{D}^{\pi}(z, \vec{q}, \vec{r}_{\perp}, \vec{b}, \mu^{2})}{ \pi z}.  \nn
\eea
In the equation above, $S_{A/B} (x_{a/b}, \vec{r}_{\perp}, \vec{b}, \mu^{2})$ represents the shadowing of nucleon PDFs in a large nucleus, which depend on the 
the impact parameter of the collision and the transverse location of the actual nucleon in the nucleus. The $\widetilde{D}$ represents the medium modified fragmentation 
function. Any and all corrections to the final fragmentation function, whether of partonic or hadronic origin, are included in this function. This depends, not only on the 
momentum fraction and the scale, but also on the momentum of the parton $\vec{q}$ as well as the origin in transverse space of the hard jet approximated by the 
location of the nucleon-nucleon collision that led to the production of this hard jet.  The overall factor $t_{AB}$ is defined as,
\bea
t_{AB} (\vec{r}_{\perp}, \vec{b}) &=& \int dz_{A} dz_{B} \rho_{A} \left(  \vec{r}_{\perp} + \frac{\vec{b}}{2} , z_{A}\right) \rho_{B} \left(  \vec{r}_{\perp} - \frac{\vec{b}}{2} , z_{B}\right).
\eea
The quantity that may be familiar to many readers, the thickness function, may be obtained as $T_{AB} (\vec{b}) = \int d^{2} r_{\perp} t_{AB} (\vec{r}_{\perp}, \vec{b})$.

In such a formalism, almost the entire calculation of final state jet modification lies within the calculation of the medium modified fragmentation function $\widetilde{D}$. 
In some schemes, such as AMY and HT, the calculation of the medium modified fragmentation function is carried out by following the full shower, where each parton scatters and splits into more partons, with virtuality dropping in the HT scheme as the shower proceeds in time. 
At the end of the shower, one folds the distribution of partons with a fragmentation functions at the 
scale of the partons.  
However in schemes such as the GLV, one focusses solely on the leading parton, and ignores the fragmentation of the radiated gluons. There is no evolution of 
shower with a dropping virtuality, and one follows the leading parton as it scatters, radiates and loses energy. 
In either methodology, the fragmentation of the hard parton (partons) takes place in the vacuum, post exit from the medium, and only the propagation of the partons 
is effected by the dense medium. As a hard parton propagates through the dense medium, it will endure multiple scatterings and 
will radiate multiple gluons, these will in turn, endure more scattering and radiate more gluons. The means to solve this problem (within or without the leading parton approximation) is to first calculate the probability or 
cross section to radiate a single hard gluon, stimulated by an arbitrary number of multiple scatterings. This single gluon emission rate is then iterated to obtain the multiple 
gluon emission rate.

\begin{figure}[!ht]
\begin{center}
\includegraphics[width=1\columnwidth]{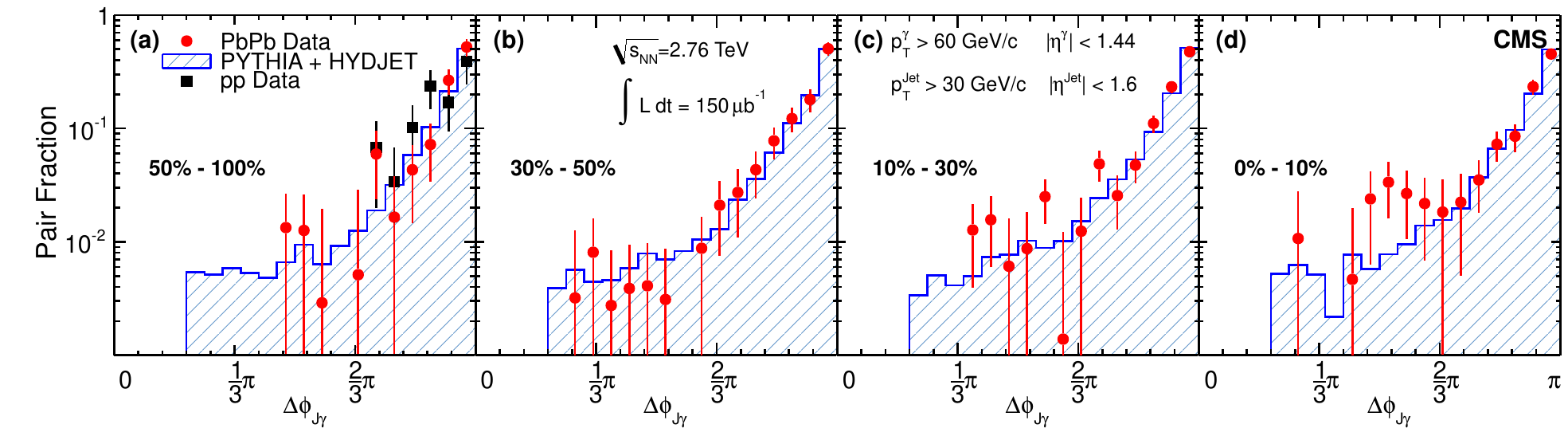}
\caption{The azimuthal distribution of jets associated with the production of a high-$p_{T}$  photon. Plot taken from the arXiv version of Ref.~\cite{Chatrchyan:2012gt}. 
Experimental data reproduced here with permission from CERN.}
\label{CMS-gamma-hadron}
\end{center}
\end{figure}

In what follows, we will outline the salient features that must be incorporated within any calculation for 
both a successful comparison with data, as well as, to pass the 
test of a \emph{tenable phenomenology}. Most of the criteria described below are 
for light flavors; to some extent heavy-quarks continue to remain a minor puzzle and will be described separately. 
Some of the features below are emergent, arising out of the experience of comparing theory calculations with data (i.e. not dictated by purely theoretical 
considerations):
\\

\noindent
i) \emph{Near Eikonal trajectories of the hard jet core}: As the jet propagates through the dense medium, it radiates a shower of softer partons. While many of these will deplete their energy and end up being stopped by the medium, for high energy jets, the hardest partons in the jet shower, tend to escape the medium and fragment outside. These hard partons 
acquire minimal 
acoplanarity in course of propagation through the dense medium. This has been clearly demonstrated by $\gamma$-jet azimuthal angular correlations, as measured by the CMS collaboration and reproduced here in Fig.~\ref{CMS-gamma-hadron}. As the reader will note, that even in the most central event, up to an angle of $\Delta \Phi \lesssim \frac{2 \pi}{3}$, there is no evidence of any excess acoplanarity over what is expected from extrapolated $p$-$p$ collisions. This would indicate that the hard parton which escapes the medium 
has either little interaction or has only soft multiple exchanges with the medium, which do not noticeably change the direction of propagation. Given that transverse momentum 
exchanges with the medium are driven by the virtuality scale, i.e. $\lc k_{\perp}^{2} \rc \sim \mu^{2}$, this suggests that the medium would appear somewhat dilute to the jet at the virtuality scale. 
\\

\noindent
ii)  \emph{Inclusion of vacuum radiation and vacuum medium interference in radiation}: Hard jets are formed in hard interactions in $p$-$p$ or $A$-$A$ collisions and 
thus these jets produce a vacuum shower of partons, even in the absence of a medium. In the presence of a medium, these showers are modified by scattering off the 
constituents in the medium. Calculations of jet modification need to include, not only radiation that is induced by the medium, but also vacuum like radiation, and the interference between 
vacuum and medium induced radiation. In fact, part of the rise in the $R_{AA}$ at very high $p_{T}$, at LHC energies, is due to this effect. Prior calculations in the 
AMY formalism were carried out in the large length limit where one ignores both vacuum like and vacuum medium interference. 
Including these effects does improve the description of data~\cite{Young:2012dv}, indicating their importance.
\\

\noindent
iii) \emph{Running of $\A_{S}$ with the hard scale of the jet}: No doubt, in any realistic description of jet modification, $\A_{S}$ appears in various locations. There is the coupling of the 
hard radiated gluon to the hard leading parton, the coupling of the exchanged gluons with the jet and with the medium, and there is the coupling in the medium in the absence of the jet. 
Here we refer, solely, to the coupling of hard jet like partons to each other and to the exchanged gluons from the medium. The scale of the jet refers to the virtuality of the hard parton 
propagating through the dense medium. If the scale of the medium is considered to be a very soft scale $T \sim \lambda^{2} E$ where  $E$ is the energy of the hard jet, $T$ is the 
temperature of the medium and $\lambda \ll 1$ is a dimensionless parameter, then the virtuality is an intermediate hard scale $\mu \sim \lambda E$. Alternatively, one may consider the 
scale furnished by $l_{\perp}$, the momentum component of the radiated gluon transverse to the hard parton's axis or jet axis (we are assuming that $l_{\perp} \sim \lambda E$, as 
$l_{\perp}$ is of the order of the virtuality). 
This dependence of the $\A_{s}$ on the hard scale of the 
virtuality, leads to a reduction in the energy loss for higher energy jets due to the waning of the coupling at larger energies. Such corrections, absent in prior versions of both the GLV and AMY schemes have now been included~\cite{Xu:2014ica,Young:2012dv}. 
\\

\noindent
iv) \emph{Prevalence of ordered emission (virtuality evolution) over unordered rare emissions}: 
The first set of complete calculations of jet energy loss were carried out by 
Baier-Dokshitzer-Mueller-Peigne-Schiff (BDMPS)~\cite{Baier:1996kr,Baier:1996sk,Baier:1998yf,Baier:2001yt}, based on the earlier work by Gyulassy and Wang~\cite{Wang:1991xy,Wang:1994fx}. 
The BDMPS calculations described a hard jet which had radiated off a large part of its virtuality in vacuum and would enter the medium with high energy close to its mass shell. 
Gluons radiated from such a parton would be de-correlated away by diffusion caused by multiple scattering. In the absence of a medium, the jet would radiate minimally. The 
energy loss from such jets tends to have an $L^{2}$ dependence. This can be easily understood by considering a hard on shell parton that radiates a gluon fluctuation, this virtual 
gluon gains transverse momentum over the length $L$ traversed by it, 
\bea
\lc p_{T}^{2} \rc = \hat{q} L.
\eea
If the radiated gluon is space like off-shell by $\mu^{2}$, it will require a $\lc p_{T}^{2}\rc \sim \mu^{2}$ to go back on shell. The gluon 
will have to gain this amount of transverse momentum within its lifetime of $\omega/\mu^{2}$, where $\omega$ is the energy of the gluon. As a result, we obtain, 
\bea
\mu^{2 } \sim \lc p_{T}^{2} \rc \sim \frac{\omega}{L}, \,\,\,\, \Rightarrow \,\,\,\, \w \sim \hat{q} L^{2}. 
\eea

\begin{figure}[!ht]
\begin{flushleft}
\includegraphics[width=0.49\columnwidth]{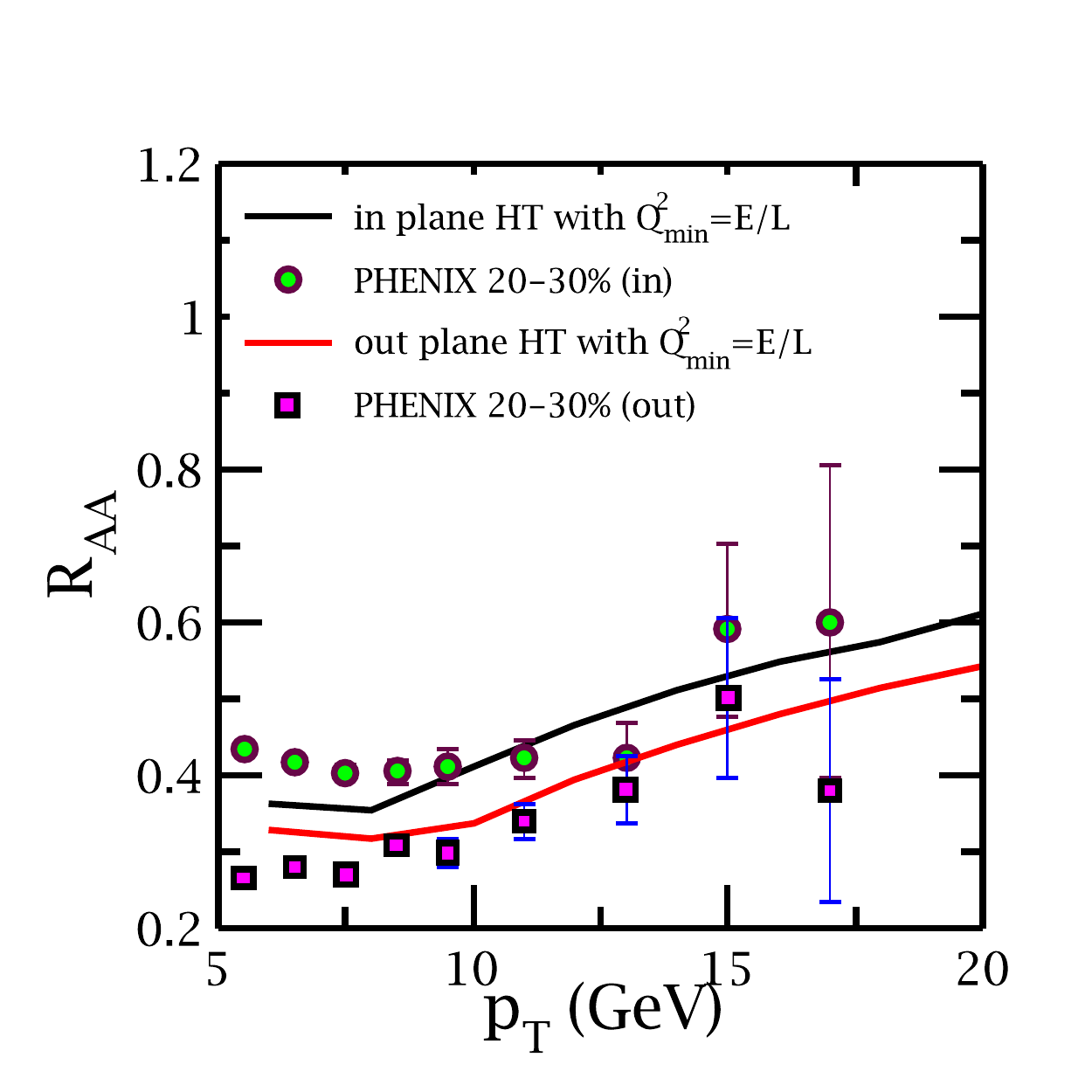}
\includegraphics[width=0.49\columnwidth]{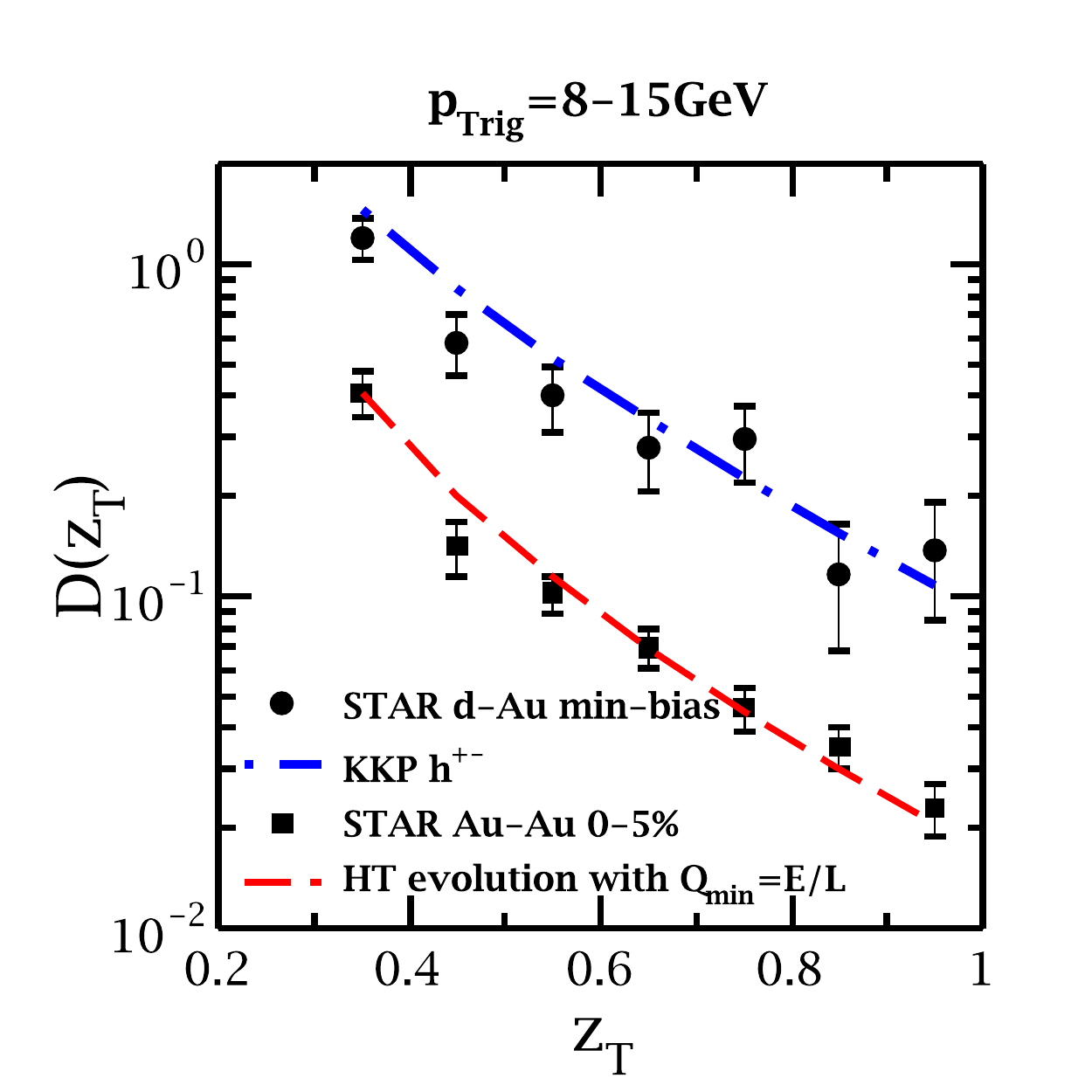}
\caption{Left plot:The in-plane and out-of-plane $R_{AA}$ at RHIC energies, calculated in the HT scheme and compared with PHENIX data for 
20-30\% central events. 
Right plot: The yield of associated hadrons per trigger as a function of the associated momentum fraction $z_{T} = p_{T}^{assoc.} /  p_{T}^{trig.}$ in 
$d$-$Au$ and 0-5\% most central collisions. All calculations are parameter free, the one parameter $\hat{q}_{0}$  is dialed to one data point in the 
$R_{AA}$ for 0-5\% most central collisions. See text for details.}
\label{PHENIX-in-out-raa-HT}
\end{flushleft}
\end{figure}

The energy lost is the gluon's energy, thus we have $\Delta E \sim \hat{q} L^{2}$.
The radiations from such a parton are rare and must be enhanced by increasing $\hat{q}$ which also increases the transverse broadening (acoplanarity) of the hard jet. 
The observation of the azimuthal asymmetry of jet quenching as a function of the azimuthal angle with respect to the reaction plane, seems to require an energy loss 
that is proportional to a slightly higher power of $L$. In the case of very virtual jets, where the jet radiates gluons and ``semi-monotonically'' loses virtuality in the medium via 
multiple ordered emissions, the 
loss of virtuality depends on the length traversed. The word ``semi-monotonically'' is used above as virtuality may both rise or drop due to scattering in the medium; however, over 
larger lengths involving multiple emissions, virtuality always tends to drop. As a result, there is a stronger length dependence of energy loss beyond $L^{2}$~\cite{Majumder:2009zu}.  
The effect of this stronger length dependence can be immediately seen in the $R_{AA}$ as a function of the angle with the reaction plane for soft particle production~\cite{Majumder:2006we}. 
Plotted in the left plot of Fig.~\ref{PHENIX-in-out-raa-HT}, is the $R_{AA}$ for high $p_{T}$ hadrons traveling parallel to the reaction plane (in-plane), 
and perpendicular to the reaction plane (out-of-plane), of a semi-central (20-30\% centrality) collision. The shorter length in-plane leads to less 
suppression than that along the longer 
length out of plane. Calculations are from the HT scheme, including virtuality evolution, 
which provides a good description of the data for $p_{T} \gtrsim 8$~GeV.

Yet another test of the length dependence of the formalism is the away side associated yields of high $p_{T}$ hadrons per trigger. This is plotted in the right hand panel of 
Fig.~\ref{PHENIX-in-out-raa-HT}. One triggers on a high $p_{T}$ hadron, in this case between 8-15 GeV, and then measures the yield in bins of 
$z_{T} = p_{T}^{assoc.} /  p_{T}^{trig.}$ on the away-side. Due to trigger bias (the tendency of detected trigger hadrons to originate close to the surface), the 
hard partons that fragment into the associated hadrons on the away side, tend to travel a longer path-length in the medium. The calculations for the medium modified 
fragmentation function, presented in the 
right hand plot, are similar to those in the left hand plot, besides the issue of longer length in the medium. Connected with the dihadron yield on the away side is the 
yield of dihadrons on the near side~\cite{Majumder:2004pt}. 
The description of such observables requires the introduction of a new non-perturbative object: 
the dihadron fragmentation function~\cite{Majumder:2004br,Majumder:2004wh}, which shows yet a different 
relationship with the medium length. The surface bias of the trigger jet at RHIC energies has led to the observation of minimal modification 
in the near side associated yield. As such, there has been less work on this topic. 
\\

\begin{figure}[!ht]
\begin{flushleft}
\includegraphics[width=0.495\columnwidth,height=0.415\columnwidth]{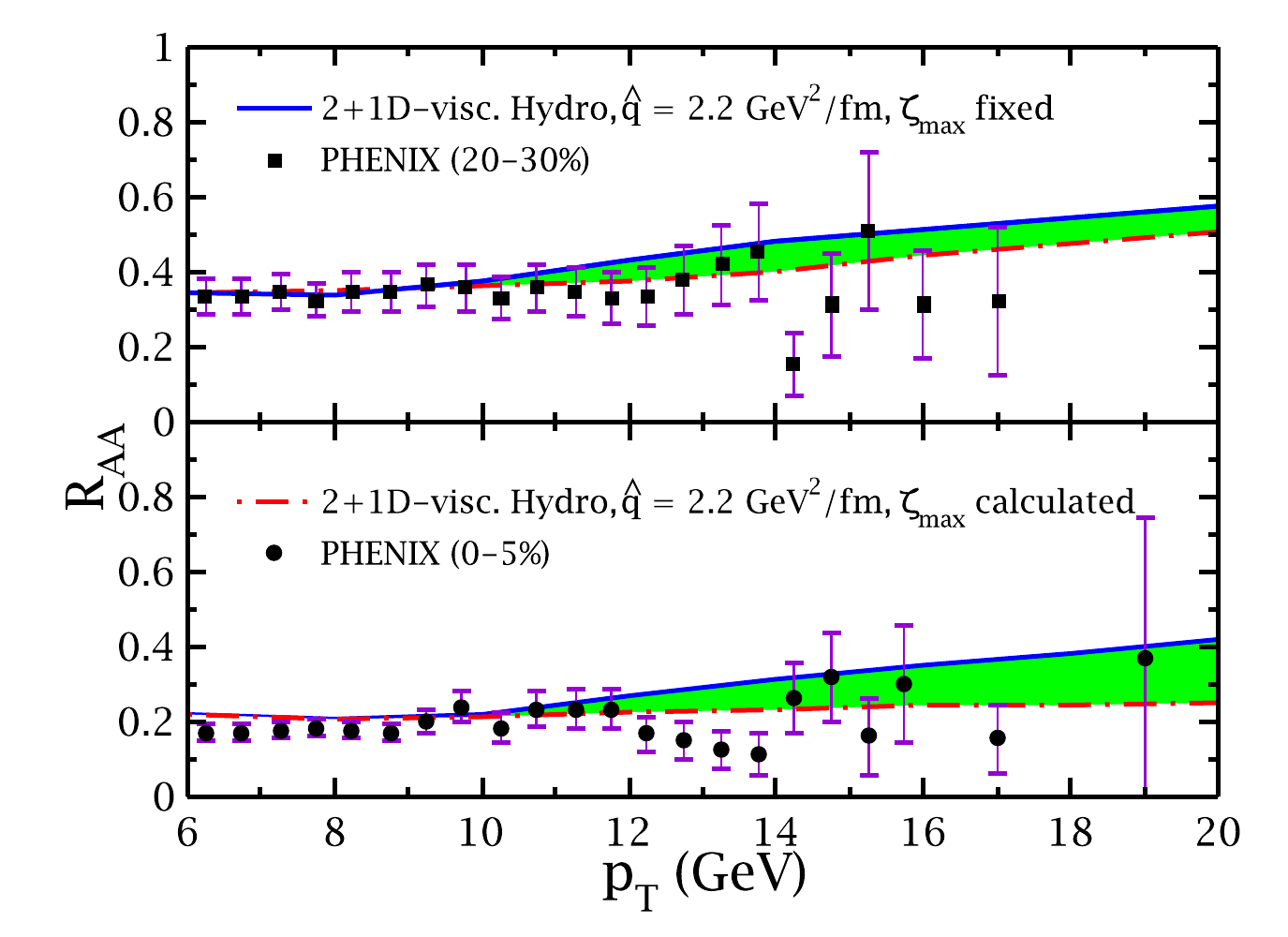}
\includegraphics[width=0.495\columnwidth,height=0.415\columnwidth]{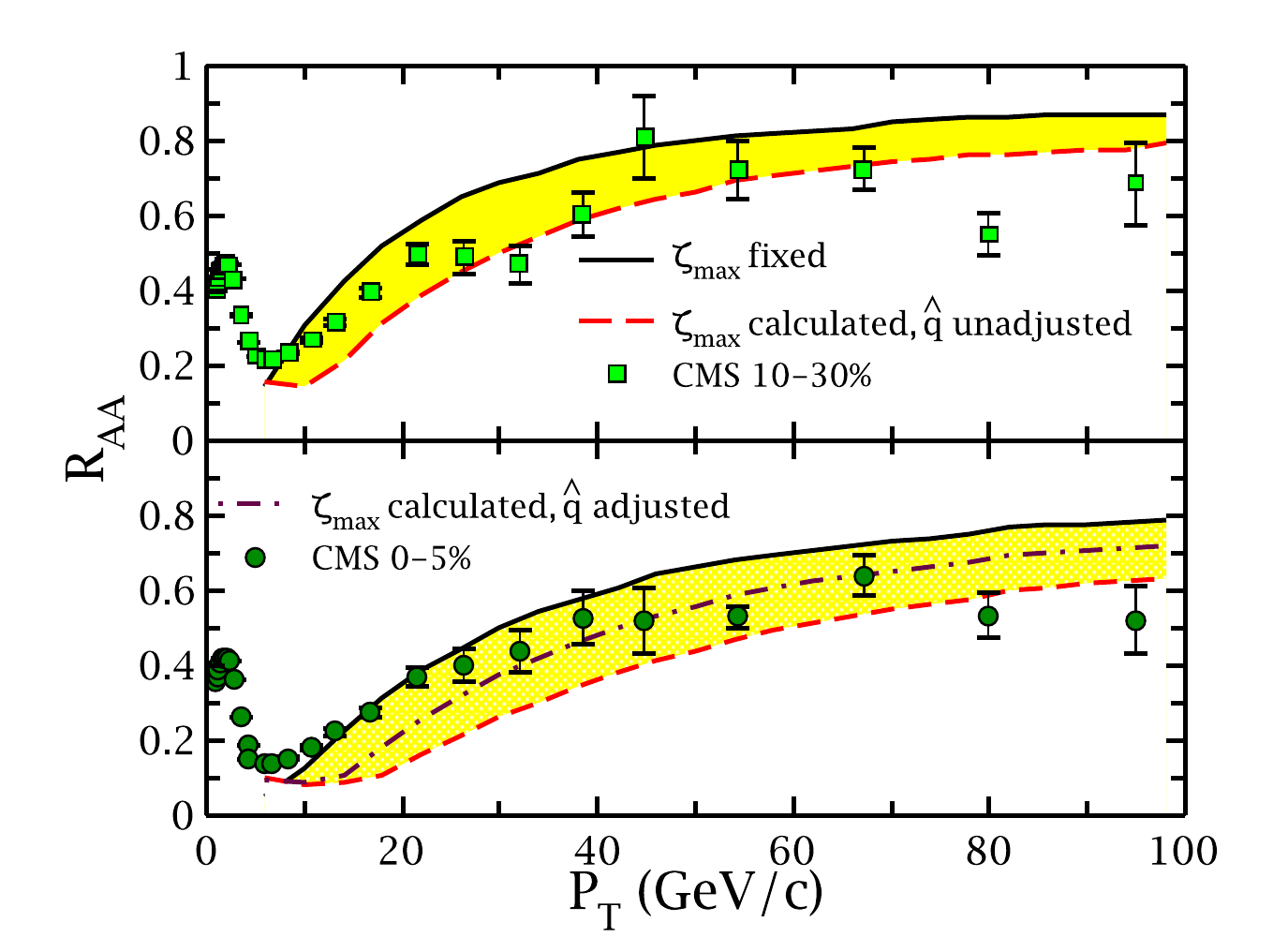}
\caption{Left plot: The $R_{AA}$ in 0-5\% central and 20-30\% semi-central events at RHIC energies, calculated using length dependent virtuality evolution codes within the HT scheme. The single parameter $\hat{q}_{0}$ is set using the data points around $p_{T} \simeq 10$ GeV for 0-5\% most central collisions.  
Right plot:  The $R_{AA}$ in 0-5\% central and 10-30\% semi-central collisions at LHC energies. Calculations are carried out, parameter free, by extrapolation from RHIC results. 
See text for details.}
\label{CMS-PHENIX-RAA-HT}
\end{flushleft}
\end{figure} 

\noindent
v) \emph{Incorporation of a realistic simulation of the medium}:  At both RHIC and LHC energies, viscous fluid dynamics simulations, either with a hadronic cascade as an afterburner, 
or with the viscous fluid dynamics continuing into the hadronic phase, have furnished a very accurate description of the data. 
If one accepts the underlying physical mechanism in these simulations, then one is compelled to also accept the space-time profile of the energy density $\epsilon$, 
the temperature $T$ and other intrinsic quantities, such as the entropy density $s$ etc. as determined by these simulations. 
Such space-time profiles become stringent constraints on the calculation of jet modification:
Can the measured jet observables be calculated if one stipulates the medium to be described by fluid dynamics? At this time of writing, many of the leading jet modification calculations have been incorporated on a viscous fluid dynamical simulations. Ref.~\cite{Bass:2008rv} represents the first instance when three different formalisms were constrained to run on the 
same hydrodynamical background. In current calculations, one has both event-by-event and event averaged simulations, where even the thickness function that controls the space-time distribution of jet production is furnished by the fluid dynamical simulation. Jet modification calculations not subject to such strict constraints often show better agreement with the full range of experimental data, due to the greater freedom available to parametrize the medium. An outstanding example of this is the $R_{AA}$ as a function of the angle $\phi$ with respect to the reaction plane, 
calculations done on a medium without hydrodynamic expansion naturally demonstrate a stronger dependence on the angle with the reaction plane~\cite{Majumder:2006we,Majumder:2007ae,Xu:2014ica}. 
\\

An example of a formalism that naturally fulfills the above requirements (i.e., without the inclusion of additive terms beyond the base formalism) is that of the Higher-Twist (HT) scheme. In this approach the medium modified 
fragmentation function is calculated via an evolution equation:  
\bea
\frac{\prt {D_q^h}(z,\mu^2\!\!,p^-)|_{\zeta_i}^{\zeta_f}}{\prt \log(\mu^2)} &=& \frac{\A_s (\mu)}{2\pi} \int\limits_z^1 \frac{dy}{y} 
P(y) \left[ {D_q^h}\left. \left(\frac{z}{y},\mu^2\!\!,p^-y\right) \right|_{\zeta_{i}}^{\zeta_f}  \right. \nn \\
&+& \left. \int\limits_{\zeta_i}^{\zeta_f} d\zeta K_{p^-,\mu^2} ( y,\zeta)  {D_q^h}\left. \left(\frac{z}{y},\mu^2\!\!,p^-y\right) \right|_{\zeta}^{\zeta_f} \right] .  \label{in_medium_evol_eqn}
\eea
In the equation above, $z$ is the momentum fraction of the fragmenting hadron, $\mu^{2}$ is the virtuality of the jet, $p^{-}$ represents the light-cone momentum of the jet along its axis.
The jet is assumed to be formed at a location $\zeta_{i}$ and exits at a location $\zeta_{f}$. The first term in the top line of Eq.~(\ref{in_medium_evol_eqn}), represents the 
vacuum evolution contribution and the term in the second line represent the in-medium contribution, where a single gluon emission is modified by multiple scattering in the medium.
The leading twist contribution to the multiple scattering, single emission kernel $K$, is given as, 
\bea
K_{p^-,\mu^2} ( y,\zeta) = \frac{2 \hat{q} }{\mu^{2}}
\left[ 2 - 2 \cos\left\{ \frac{\mu^2 (\zeta - \zeta_i)}{ 2 p^- y (1- y)} \right\}  \right] . \label{kernel}
\eea
In the evaluation of the above expression, we have also carried out a Taylor expansion of the kernel in ${k_{\perp}^{2}}/{\mu^{2}}$, and retained the 
leading and next to leading corrections~\cite{Majumder:2009zu}. 

The one unevaluated parameter in the expression above is the  jet transport coefficient $\hat{q}$. This represents the soft matrix elements which have been factorized out 
from the hard part of the energy loss calculation. This admits a gauge invariant operator expression~\cite{Benzke:2012sz},

\bea
\mbx\!\!\!\!\!\hat{q} &=& \frac{ 4 \pi^{2} \A_{s} }{N_{c}}  \int \frac{ d y^{-} d^{2} y_{\perp} d^{2} k_{\perp}}{(2\pi)^{3}} e^{i \frac{k_{\perp}^{2} y^{-}}{2 q^{-}} 
-i k_{\perp} \cdot y_{\perp} }  \label{qhat_gaugeinvar} \\
&\times& \left\langle P \left| \tr \left[ 
U^{\dag}(\infty^{-}, 0_{\perp} ; 0^{-}, 0_{\perp} ) \right. \right. \right. 
T^{\dag}(\infty^{-},\vec{\infty}_{\perp};\infty^{-},0_{\perp})
 \nn \\ 
\ata \left. \left. \left. T(\infty^{-},\infty_{\perp};\infty^{-},y_{\perp}) U(\infty^{-}, y_{\perp} ; y^{-}, y_{\perp} ) 
t^{a} {F_{\perp}^{a}}^{+ \mu} (y^{-}, y_{\perp})
\right. \right. \right. \nn \\ 
\ata U^{\dag}(-\infty^{-}, y_{\perp} ; y^{-}, y_{\perp} )  T^{\dag}(-\infty^{-},\vec{\infty}_{\perp};-\infty^{-},y_{\perp})
 \nn \\ 
\ata \left. \left. \left. T(-\infty^{-},\infty_{\perp};-\infty^{-},0_{\perp})
U^{\dag}(0^{-}, 0_{\perp} ; -\infty^{-}, 0_{\perp} ) t^{b}{ F_{\perp}^{b}}^{+}_{,\mu}(0^{-},0_{\perp}) \right] \right| P \right\rangle . \nn
\eea
In the equation above, ${F_{\perp}^{a}}^{+ \mu} (y^{-}, y_{\perp})$ represents the gluon field strength, $t^{a}$ is a Gell-mann matrix. 
The factors $U(x^{-},\vec{y}_{\perp};y^{-},\vec{y}_{\perp})$ and $U^{\dag}(x^{-},\vec{y}_{\perp};y^{-},\vec{y}_{\perp})$ represent gauge links along 
the negative light-cone direction. Along with these we also have the transverse gauge links at negative light-cone $\pm \infty$: 
$T(\infty^{-},\infty_{\perp};\infty^{-},y_{\perp}) $ and $T^{\dag}(\infty^{-},\infty_{\perp};\infty^{-},y_{\perp}) $. The 
expectation of the operator is obtained in the state $| P \rc$. This may either represent a single state as that in a large nucleus, or could be 
replaced with an ensemble average with a particular density to represent a finite temperature environment.

While the general, gauge-invariant expression given above for $\hat{q}$ is rather complicated, in a particular gauge, such as $A^{\pm} = 0$ gauge, this may 
become much more simplified. Since $\hat{q}$ represents the soft matrix element in a hard process which is evaluated using pQCD, a gauge 
has to be picked prior to any calculation. Once a gauge is picked and $\hat{q}$ factorized as in Eq.~\ref{kernel}, one still has to evaluate it. One 
option is to scale it with a similarly dimensioned intrinsic property of the medium, such as $s$, $\epsilon^{3/4}$ or $T^{3}$, e.g.,
\bea
\hat{q} (\vec{x},\tau) = \frac{\hat{q}_{0}}{s_{0}} s(\vec{x},\tau). \label{qhat-scaling}
\eea

 Another avenue that may become available in the near future 
is to calculate $\hat{q}$ on the lattice~\cite{Majumder:2012sh,Panero:2013pla}. The former method has been used so far, with experimental data 
being used to determine $\hat{q}_{0}$ as the maximum value of $\hat{q}$, in the center of the most central collision at $\tau = 0.6$fm/c, at top 
RHIC energy of $\sqrt{s} = 200$GeV/nucleon-pair. Once set this way, the value of $\hat{q}$ at any location, in any centrality, at any energy of 
collision may be determined by using the formula above and a knowledge of the entropy density at the location in question. 
The equation for the relation of $\hat{q}$ with the local entropy density $s$, given above, is valid only at very high temperatures, for a static media. 
While corrections for local boost of the medium can be straightforwardly introduced, there is still a lack of knowledge, regarding the variation of 
$\hat{q}$ with $T$ (and/or $s$) near and below the transition temperature $T_{C}$. This will be discussed further in Sec.~\ref{new-stuff}.

Incorporating the set of constraints elucidated above, assuming the relation between $\hat{q}$ and the local $T$ (or $s$), and setting $\hat{q}_{0}$ to fit 
the $10$GeV data point in 0-5\% central collisions at RHIC, one may obtain a reasonable description of the $p_{T}$, centrality and energy dependence of 
the single hadron $R_{AA}$. This is plotted in Fig.~\ref{CMS-PHENIX-RAA-HT}: the left panel shows the fit at RHIC energies and the right hand plot shows the
parameter free extrapolation to LHC energies. The more differential calculations of $R_{AA}$ versus reaction plane and the away side associated yield were 
presented in Fig.~\ref{PHENIX-in-out-raa-HT}.

In this section, we have outlined, the basic ingredients that must be included within a jet energy loss formalism to be successful at describing RHIC and LHC 
data on few particle observables. Prior to the extension to jet observables, any scheme of energy loss has to fulfill the criteria mentioned above.
In the subsequent sections we will discuss the extension to full jet observables followed by a discussion of the new directions for theoretical research in 
jet modification.

\section{Full Jet Observables}~\label{full-jet}

With the order of magnitude increase in center-of-mass energies at LHC compared to RHIC, the jet production cross section for jets above 20GeV has increased dramatically.
While full jet modification measurements were first carried out at RHIC, the energy reach of the LHC has allowed the LHC heavy-ion experiments to carry out much more differential jet modification studies, at the full jet level, with greater precision. 
Jet reconstruction algorithms, nowadays, are routinely used to sequentially pair particles close in angular space into one overall four vector and then extract these jets 
event-by-event from a fluctuating background. While several reconstruction algorithms are available, the requirements of infrared and collinear safety as well as speed of use 
have narrowed down this choice to the anti-$k_{t}$ algorithm~\cite{Cacciari:2008gn,Cacciari:2008gp}. 
This algorithm codified in the FASTJET package~\cite{Cacciari:2011ma} has now become almost the universal standard in jet reconstruction 
used in both $p$-$p$ and heavy-ion collisions. In this article, we will not discuss the various possible choices of the jet reconstruction algorithms; instead we will focus on the 
physics that can be harvested from the reconstructed jets.

With the incorporation of a reconstruction algorithm, the method of calculation of jet modification has also changed. Event averaged calculations involving few particle 
inclusive fragmentation functions must be replaced by Monte-Carlo (MC) based event generators. These event generators calculate the modification to the entire distribution of 
partons in a hard jet on an event-by-event basis, as they propagate through a dense medium, which is also generated event-by-event. One keeps track of the space-time 
and momentum information of all the partons. Many of these may lose energy and become completely absorbed in the medium; some of these will have deposited sufficient 
energy within the medium to set up an excited ``wake'' that follows the path of the jet; and some part of the shower may escape the medium and hadronize externally in vacuum.
A jet reconstruction algorithm such as FASTJET will recombine particles from all these sources, as well as particles from the bulk which are uncorrelated with the jet, as part of 
a jet. As a result, all particles in an event are recombined within a finite number of jets. One may then subtract out the average $p_{T}$,  that originates from the uncorrelated background (experiments also carry out an unfolding procedure to remove the effect of background fluctuations),  
to obtain the $p_{T}$ distribution and other observables involving the reconstructed jets. Each experimental collaboration has a slightly different method of background 
subtraction and we will not discuss this further here. However, given the general format of this subtraction, the obtained jets include not only the surviving 
hard parton (partons), fragmenting in vacuum, but also hard collinear partons stimulated by scattering in the medium and the ``wake'' of deposited energy that follows the jet. 

 There are two general categories of MC event generators that are currently available. \emph{Top-down} generators, 
 which start from the analytical expressions for single gluon emission and construct a probabilistic interpretation,
 which can then be sampled to obtain the medium modified shower in a given event, e.g., 
 PYQUEN/HYDJET\cite{Lokhtin:2005px,Lokhtin:2008xi}, 
 Q-PYTHIA~\cite{Armesto:2009fj}, 
 MARTINI~\cite{Schenke:2009gb,Young:2011ug} and 
 MATTER\cite{Majumder:2013re}. The other category are the \emph{bottom-up} generators which 
 start with a vacuum parton shower and then modify the matrix elements to mimic the quantum interference effects in multiple scattering stimulated emission. The modification is tuned 
 such that the final result will either match the results of 
 analytic calculations of leading parton energy loss e.g., in JEWEL~\cite{Zapp:2008af,Zapp:2008gi,Zapp:2011ya}, 
 or fit the experimental data for a single observable such as the $R_{AA}$, e.g. YAJEM~\cite{Renk:2010qx,Renk:2010zx}. 
 Once so tuned, these are then used to compute other observables. 

\begin{figure}[!ht]
\begin{flushleft}
\mbx\hspace{0.5cm}\includegraphics[width=0.45\columnwidth]{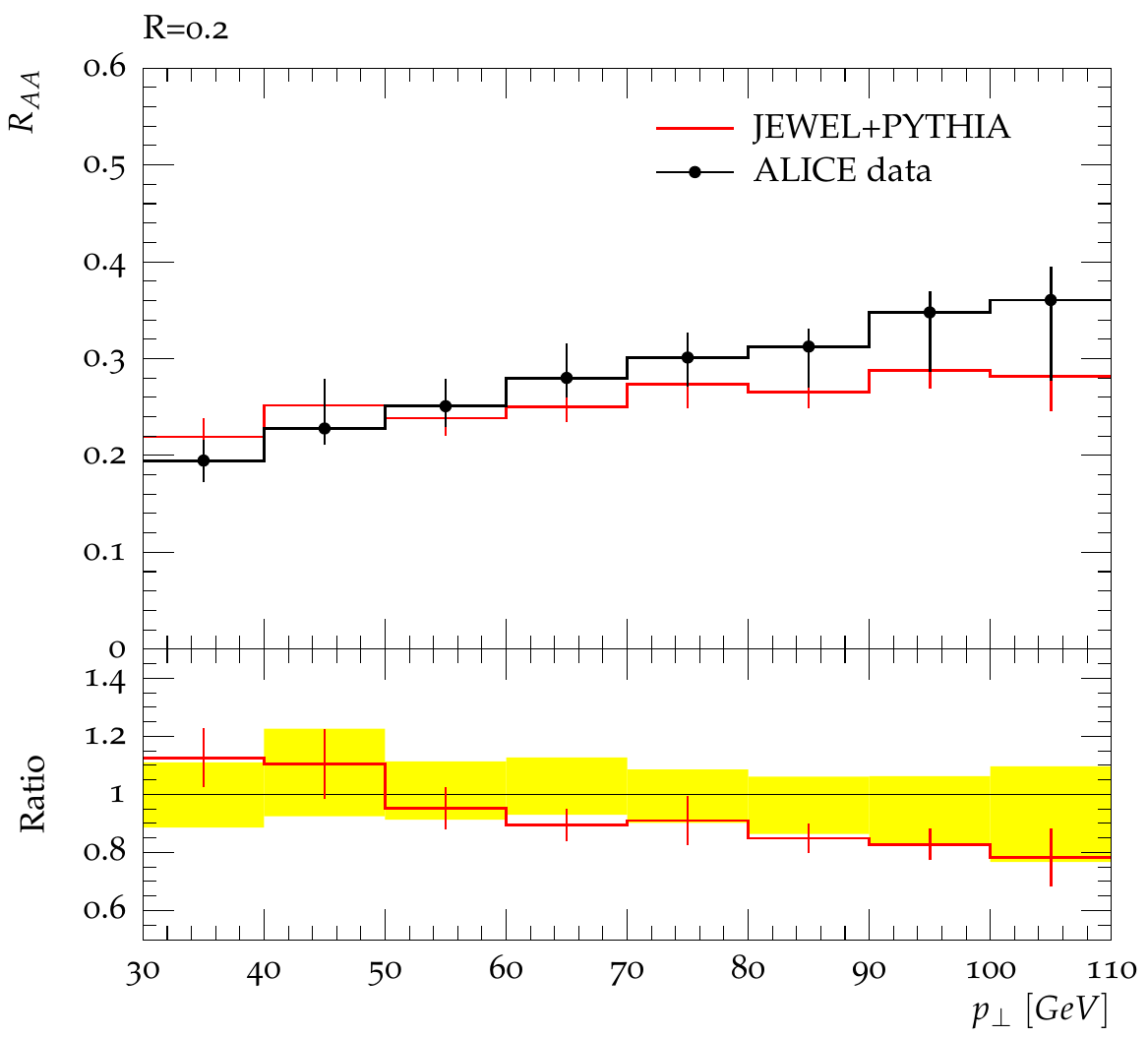}\hspace{0.5cm}
\includegraphics[width=0.45\columnwidth]{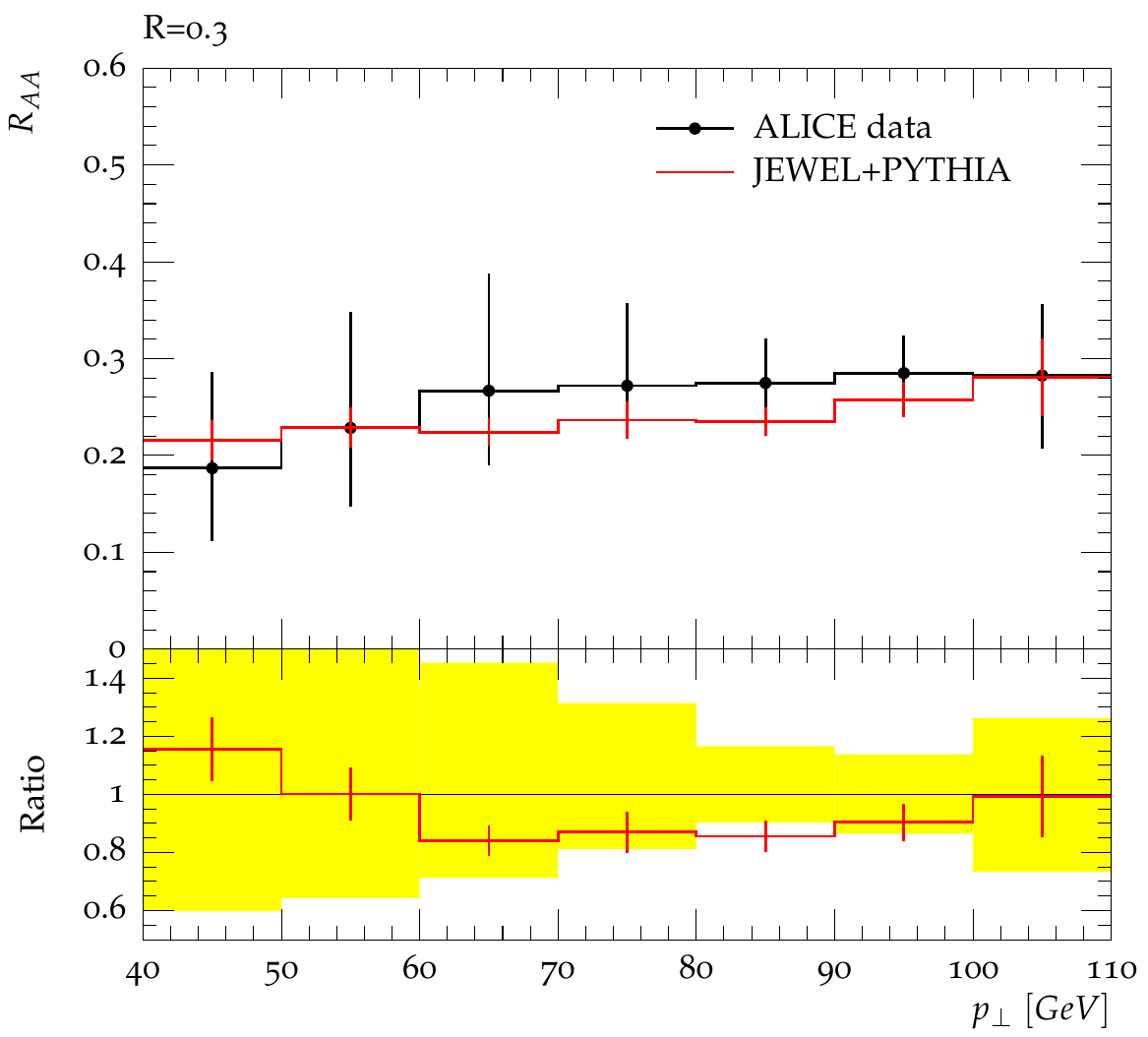}
\caption{Reconstructed jet $R_{AA}$ for two different jet resolution parameters $R=0.2$ (left plot), and $R=0.3$ (right plot). Theory calculations are from the JEWEL 
event generator~\cite{Zapp:2012ak}. Plot reproduced from the arXiv version of Ref.~\cite{Zapp:2012ak} with permission. Experimental data points are taken from 
the ALICE detector at CERN.}
\label{CMS-JET-RAA-JEWEL}
\end{flushleft}
\end{figure}

So far, full jet reconstruction measurements have focussed on four major observables: The jet $R_{AA}$, the di-jet energy asymmetry $A_{J}$ and angular acoplanarity, the fragmentation function within a reconstructed jet and the intra-jet shape. The calculation of such observables proceeds in a manner similar to the experiment, with a part of the produced shower 
reconstructed into a final jet using a jet reconstruction algorithm. As mentioned above, in actual measurements, there are in principle, three separate contributions within a reconstructed jet: 
The part of the jet that survives the medium and fragments in the vacuum; the part of the medium which has interacted with the jet, has been excited into a ``jet-wake'' by the deposition of energy and 
momentum from the jet, or become correlated by re-combinatoric hadronization with a portion of the jet (as well as hard collinear emissions from the jet that were stimulated by the medium); and 
finally the portion of the uncorrelated medium which has never interacted with the jet, yet appears within the reconstructed jet by virtue of location 
in angular space. As a result, any observable constructed using particles within a reconstructed jet, e.g., the total $p_{T}$ has three contributions, 
\bea
p_{T} = p_{T}^{vacuum-jet} + p_{T}^{jet-medium} + p_{T}^{background}. \label{reconstruction-contributions}
\eea
The background subtractions carried out by most experimental collaborations are meant to remove this last component. 
Assuming that the result of this subtraction is a faithful representation of the actual correlated portion of the reconstructed jet, theory simulations 
have tended to ignore the last component of the uncorrelated background. 
Hence, what is referred to as a reconstructed 
jet, in a theory simulation, is a combination of the part of the jet escaping the medium and fragmenting in vacuum and hard collinear emissions stimulated in the medium. 
Most current MC event generators do not include a calculation of the contribution from the deposited energy, which has not equilibrated with the medium, 
and is correlated with the 
jet direction. In terms of the equation above, the reconstructed jet $p_{T}$ in current theory calculations includes $p_{T}^{vacuum-jet}$ and a portion of $p_{T}^{jet-medium}$.  
We should mention in passing that there is an effort underway to estimate the contribution from the correlated energy ~\cite{Qin:2009uh,Renk:2013pua} (there are also attempts using a linear Boltzmann transport simulation~\cite{Wang:2014hla}).  Along with such a calculation, there is the associated problem of the fate of the lost energy. This is the energy 
deposited by a hard jet in a dense medium, which has flowed out of the jet cone. Assuming that the energy is thermalized swiftly, this maybe carried out at large angle by the 
fluid medium. There have been a series of estimations of such effects, though not in an event-by-event setting~\cite{Renk:2005si,Chaudhuri:2005vc,Qin:2009uh}. 
Incorporation of the distribution of energy outside the jet cone in MC event generators is still considered as a second order effect, and so far ignored in most simulations.

\begin{figure}[!ht]
\begin{flushleft}
\mbx\hspace{-0cm}\includegraphics[width=0.51\columnwidth]{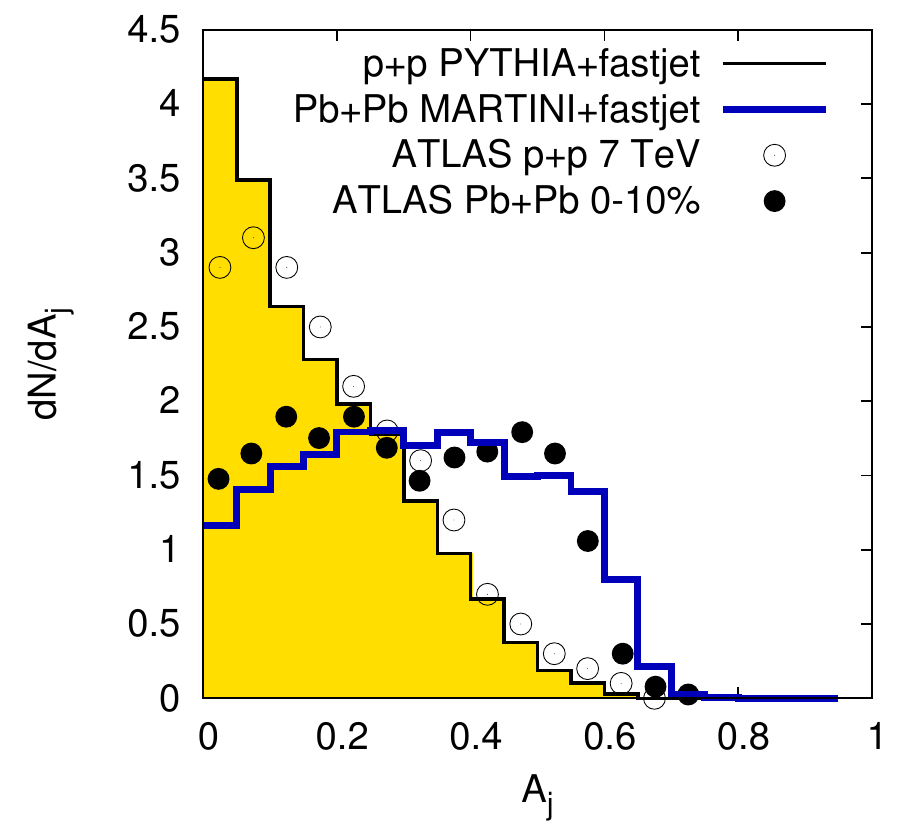}\hspace{-0.5cm}
\includegraphics[width=0.523\columnwidth]{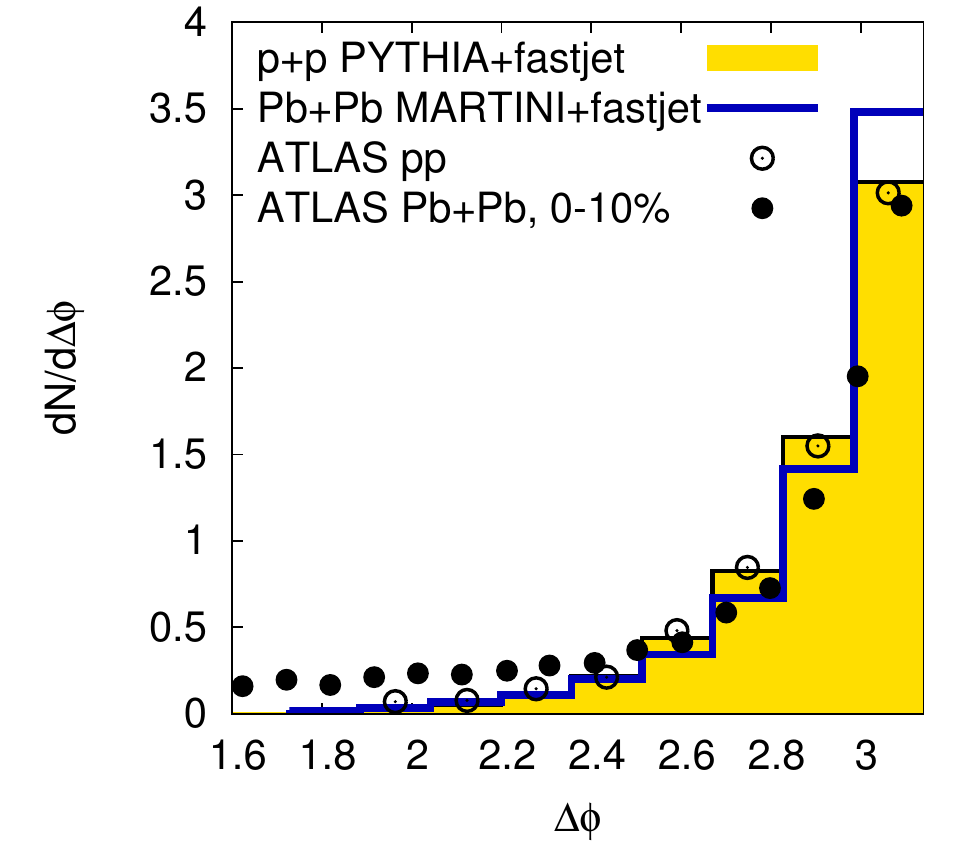}
\caption{ The distribution of the di-jet energy asymmetry $dN/dA_{J}$ (left plot), and the angular distribution of the away side jet $dN/d\Delta\phi$. Plots reproduced from 
the arXiv. version of Ref.~\cite{Young:2011va}, with permission. Data are from the ATLAS detector at CERN.}
\label{ATLAS-AJ-MARTINI}
\end{flushleft}
\end{figure}

In current MC event-by-event simulations, one no longer calculates the 
medium modified fragmentation function, instead one simulates the momentum and space-time locations of every particle in the entire shower. There are several methods of 
carrying this out, which depend to some extent on the base formalism. For example, in the HT approach, one takes the entire splitting function, including  
the vacuum and medium induced parts and constructs a Sudakov form factor, which is then sampled to obtain a virtuality ordered shower. For the case of 
MARTINI, one starts with the vacuum shower obtained from PYTHIA, which contains a distribution of partons at a low virtuality. This distribution is changed by the AMY rate equations. 
In YaJEM, one directly modifies the PYTHIA vacuum shower by introducing virtuality increases to represent the effect of the medium. In JEWEL, one follows a similar 
methodology of modifying the PYTHIA shower, with somewhat different modifications which are set up to reproduce results for the mean energy lost by the leading particle, in 
the soft scattering limit, in the BDMPS scheme.

In spite of the different approaches and the shortcomings of reconstruction discussed above, there are now several comparisons to various full jet observables.
Most successful comparisons have been to the jet $R_{AA}$ and the 
energy asymmetry observable $A_{J}$. Both these observables are 
mostly sensitive to the energy reconstructed within the entire jet cone. As a jet progresses through a dense medium, the vacuum shower, that accompanies a hard 
jet, produced in a hard interaction, is modified by the scattering in a dense medium. This scattering tends to broaden out the radiated gluons as well as stimulate more radiation 
from the hard parton. Some part of this radiation remains within the cone, either as a hard parton or as correlated deposited energy, and some part of this tends to escape the cone. 
Hence, the overall effect is that multiple scattering in a medium will tend to draw energy out of this cone, and as 
a result, a jet with an energy $E$ within a given cone, in vacuum, will find itself with energy $E - \Delta E$ after traversal through a medium. As a result, for smaller cone sizes, one  
expects an energy or $p_{T}$ dependent suppression. As one increases the energy of a jet, the radiation 
from the jet tends to become more collinear and thus is included within a narrower cone in vacuum. As a result for larger cones, one will see a reduction in the energy dependence 
of the suppression. This observable is not very sensitive to the reorganization of the energy within the jet cone itself. Several approaches have been able to describe this observable, 
we present the results from the JEWEL scheme in Fig.~\ref{CMS-JET-RAA-JEWEL}. As expected, one sees a rising $R_{AA}$ for a cone of $R=0.2$ and a rising followed by a 
flattening of the $R_{AA}$ for $R=0.3$. 
One should note that this intrinsic energy dependence is folded with the energy dependence of the falling spectrum of hard jets. Thus one 
cannot simply read off the energy dependence from the plots.

Another observable in the same vein is the fraction of events with a di-jet energy asymmetry observable $A_{J}$, which is defined as,
\bea
A_{J} = \frac{E_{T} - E_{A}}{E_{T} + E_{A}},
\eea
where, $E_{T}$ represents the energy of a reconstructed leading trigger jet and $E_{A}$ is the energy of the reconstructed associated away-side jet. Both are reconstructed using 
some resolution parameter. 

\begin{figure}[!ht]
\begin{center}
\includegraphics[width=0.85\columnwidth]{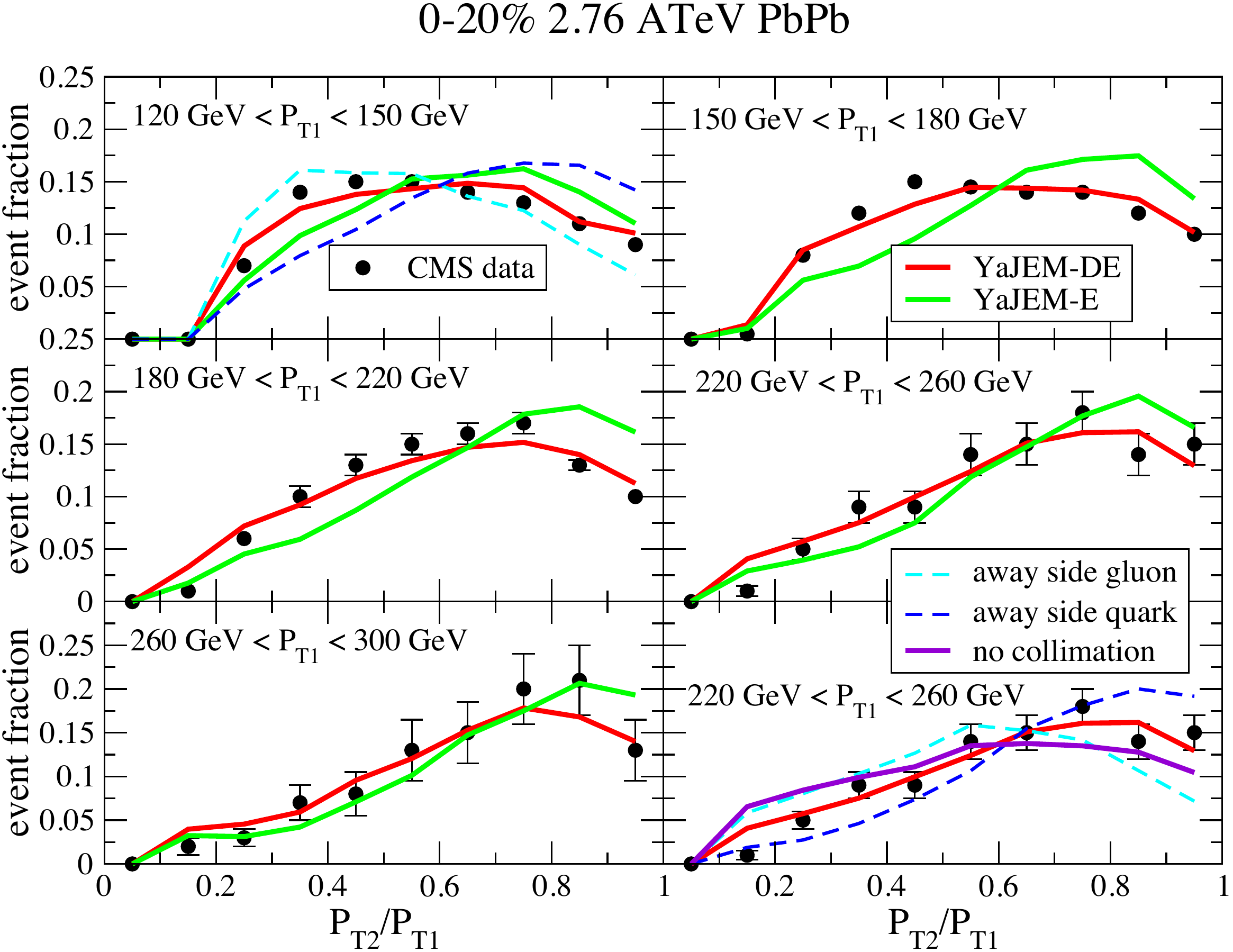}
\caption{The fraction of events with a given di-jet energy imbalance, for different jet trigger $p_{T}$, as obtained by the CMS detector in central $Pb$-$Pb$ events and 
calculated in the YaJEM event generator.  The green line includes only drag effects, while the red line includes both drag and radiative loss. Figure reproduced with 
permission from the arXiv version of Ref.~\cite{Renk:2012cb}. }
\label{renk-Aj}
\end{center}
\end{figure}

Unlike the jet $R_{AA}$, this observable is less sensitive to the incoming hard cross section. 
However, it too is not sensitive to the change in the internal momentum distribution for wide cones. 
Several schemes have been able to describe this observable as well, 
we present here results from the MARTINI event generator~\cite{Young:2011qx,Young:2011va} 
in Fig.~\ref{ATLAS-AJ-MARTINI}. 
It is interesting to note that one may obtain a good description of this data, even without the inclusion of 
effects such as finite-size corrections necessary for a successful description of the $R_{AA}$ for identified particles. 
The right hand plot in Fig.~\ref{ATLAS-AJ-MARTINI} contains the angular distribution of the away-side jet. This is consistent with the experimental 
results in Fig.~\ref{CMS-gamma-hadron}, in that there is no appreciable increase in the acoplanarity of the di-jet pair, even though there is a considerable energy loss seen in one of the 
jets in the $A_{J}$ distribution. Similar agreement has also been found using the YaJEM event generator~\cite{Renk:2012cb,Renk:2012cx,Renk:2012ve} 
which is also successful in describing both the single particle $R_{AA}$, 
$I_{AA}$ and the jet $R_{AA}$. In Fig.~\ref{renk-Aj}, a closely related quantity, the event yield for a given ratio of trigger and associated jet $p_{T}$ is plotted, for a range of 
trigger $p_{T}$'s. The findings from both these and other simulations is that a proper description of the energy imbalance requires the inclusion of both radiative and drag loss on the 
jets. The di-jet energy imbalance has been described by several formalisms, in principle, one may approximately obtain these even without an event generator, 
in fact both the $R_{AA}$ and the $A_{J}$ distributions have been obtained without a full MC simulation, see e.g., Ref.~\cite{Qin:2010mn}. 

\begin{figure}[!ht]
\begin{flushleft}
\mbx\hspace{-0.5cm}\includegraphics[width=0.51\columnwidth]{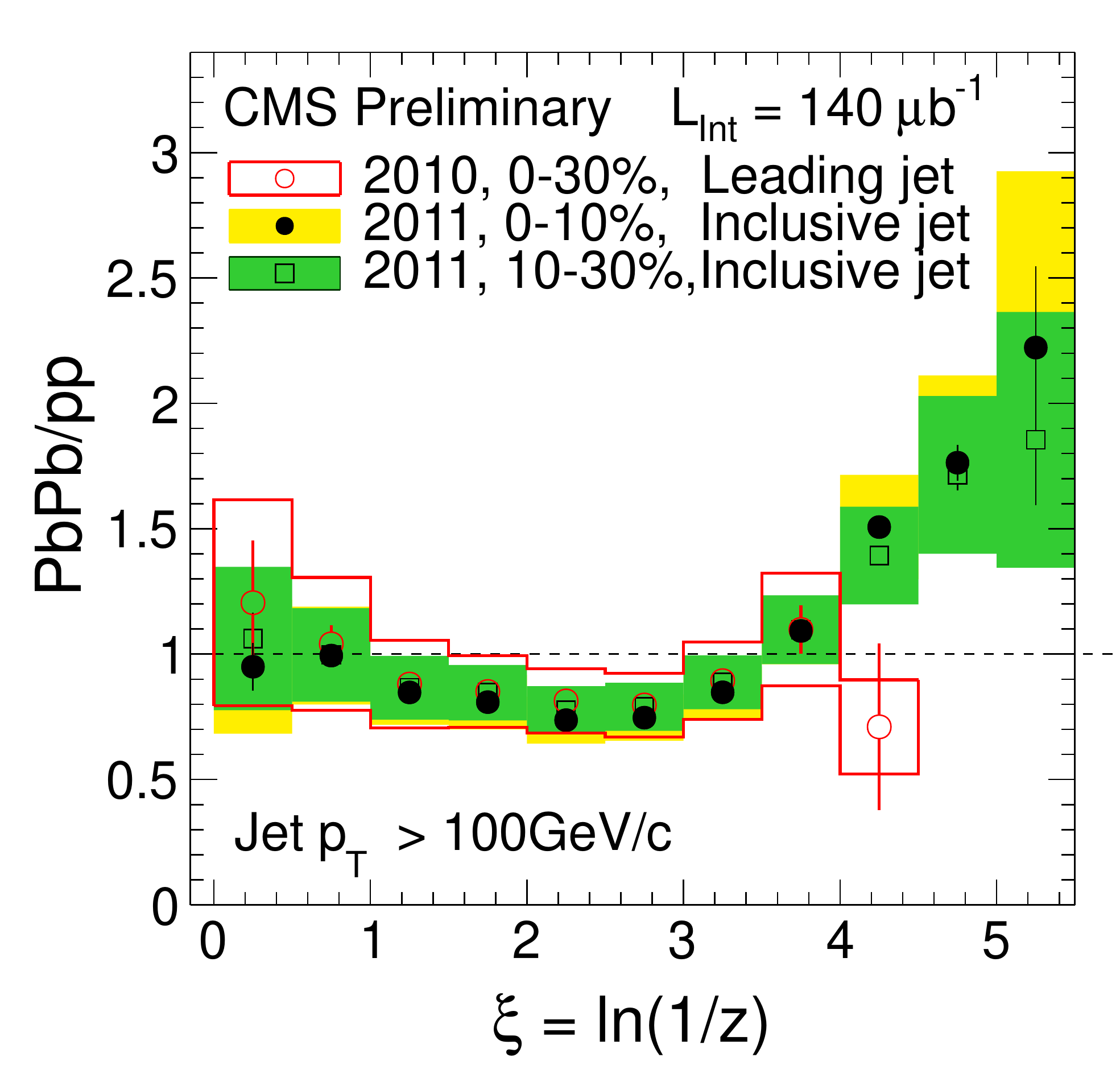}\hspace{-0.15cm}
\includegraphics[width=0.535\columnwidth]{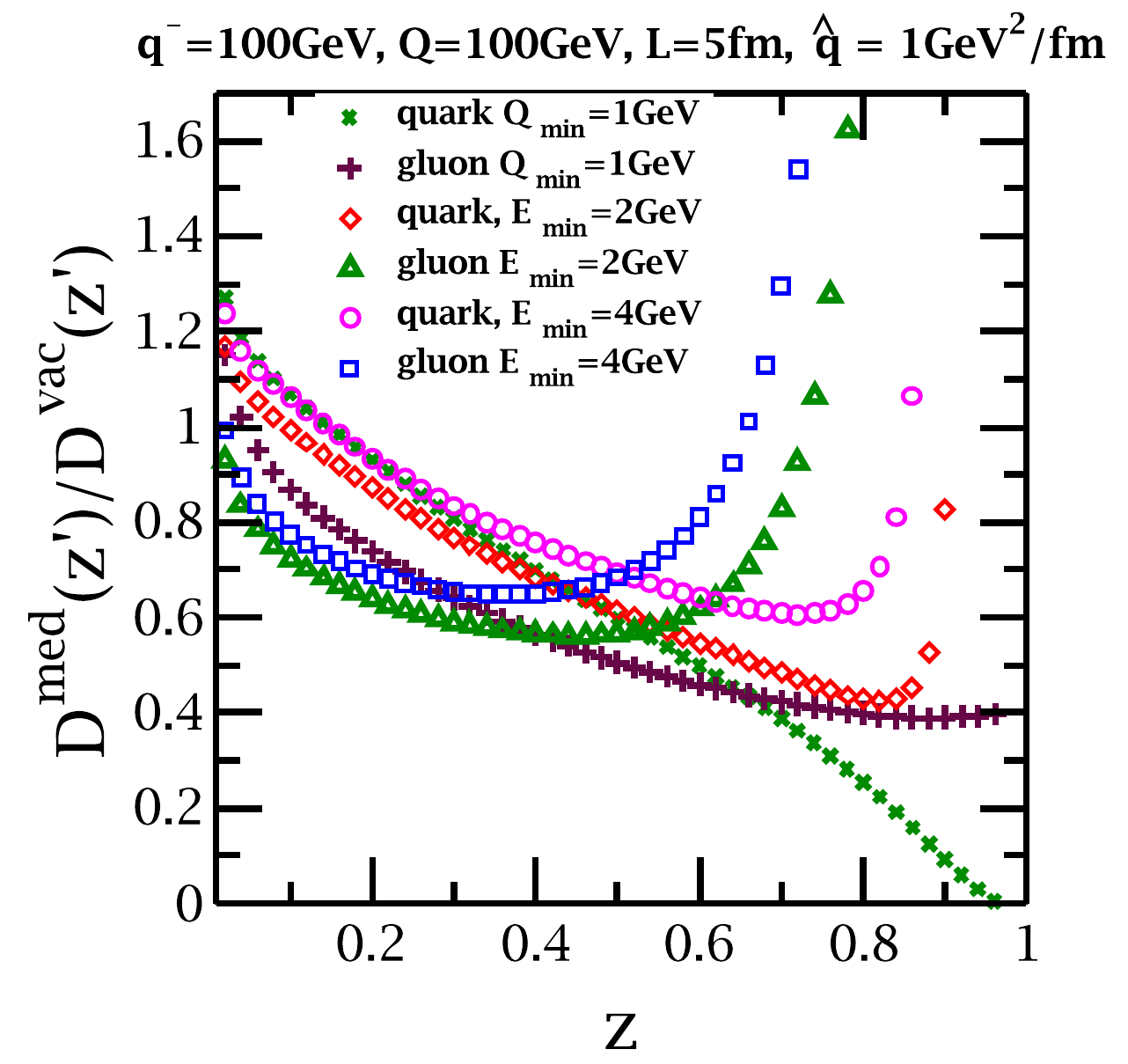}
\caption{Left: The ratio of intra-jet fragmentation function for a reconstructed jet in $Pb$-$Pb$ collisions versus that in $p$-$p$ collisions at the 
LHC, as measured by the CMS detector~\cite{CMS-PAS-HIN-12-013}. Right: A calculation of the intra-jet fragmentation function for a jet traversing a static QGP at a fixed temperature versus that 
for a vacuum jet, as calculated from the MATTER event generator~\cite{Majumder:2013re}. }
\label{CMS-frag-func-HT}
\end{flushleft}
\end{figure}

To understand the change in the internal structure of a hard jet as it passes through dense matter, one has to look at even more differential observables which pertain to the 
energy redistribution within a jet, such as the 
fragmentation function within a reconstructed jet and the jet shape which yields the energy in angular bins around the jet axis. These observables require a calculation of the 
momentum and virtuality of the hard parton (partons) as they leave the medium, as well as a calculation of the deposited energy which has not thermalized with the medium that 
shows up within the jet cone. Even though, one of the main goals of shifting from few particle observables to full jets was to eradicate the effect of hadronization, the actual simulation 
of any of these observables requires a controlled calculation of the hadronization of hard jets, both in the vacuum and in combination with the medium, along with a calculation of the 
hadronization of the excited part of the medium which forms the wake of the jet. At the time of this writing, the hadronization of parts of the jet in the presence of a medium is an 
unsettled problem. As a result, the calculation of the fragmentation function (or jet shapes) has become somewhat difficult. The experimental results from the CMS detector 
for the ratio of the fragmentation function in $Pb$-$Pb$ versus that in $p$-$p$ are plotted in the left panel in Fig.~\ref{CMS-frag-func-HT}. The experimental fragmentation 
functions are plotted in the variable $\xi = \log{1/z}$. In terms of the standard momentum fraction $z$ one sees an enhancement at $z \ra 0$ as well as at $z \ra 1$ with a 
dip at intermediate $z$. The upturn in the ratio has been described as a challenge for theoretical calculations, however, there are several effects that can cause such an effect at 
large $z$, these include a loss of some portions of the jet to the medium, due to energy loss, as well as a smaller value of virtuality for jets propagating in a dense medium compared to 
those formed in a hard collision. The first of these effects can be illustrated with a theoretical calculation in a static QGP at fixed temperature using the MATTER event generator 
in the right panel of Fig.~\ref{CMS-frag-func-HT}. These calculations clearly show, that even the loss of a small amount of energy from the reconstructed jet escaping a dense medium, 
will lead to a modification of the ratio of fragmentation functions, due to the shift in the effective $z$-variable.

While the yield at large $z$ may be predominantly expected to arrive from the fragmentation of the escaping jet, the soft $z$ part contains all the contributions highlighted in 
Eq.~\eqref{reconstruction-contributions}. As a result, the calculation of this portion requires the introduction of several aspects of the theory within the event generator, which 
are not yet straightforwardly implementable. 
The calculation presented on the right side of Fig.~\ref{CMS-frag-func-HT} is meant to be more accurate at larger values of $z$. At small $z$, all 
medium components are absent and thus, one will note that the excess is less than that seen in the data.

\section{Upcoming Developments}~\label{new-stuff}

Jet modification in dense matter is currently at a crossroads. With the advent of detailed and discerning data, several of the mechanisms
proposed to describe the quenching of hard jets have been ruled out. There is now a growing consensus on the major ingredients that must 
be included in any calculation of jet modification. Some of these were highlighted in Sec.~\ref{1-particle}. Based on these, there is now rapid development 
in the theory of jet modification on five major fronts. \\

\emph{i) MC event generator development:}  As discussed in the preceding section, there are now several observables that include the reconstruction of a full jet. 
While some may be calculated without recourse to event generators, some observables, such as the modification to the fragmentation function and the jet shape, 
require event generators that include both the change to the hard sector of the propagating jet, as well as the contribution from the wake in the medium induced by the hard jet. 
Such developments are currently underway. The major hurdle in such an undertaking is understanding the rate at which the deposited energy thermalizes in the medium 
as well as the hadronization of the jet and the medium in tandem. 
The first problem involves energy scales that are intermediate between the hard scale of the jet and the soft scale of the medium. Hence these will involve, potentially new semi-hard transport 
coefficients that deal with the rate at which hard modes are thermalized in the medium. The second issue, requires detailed modeling of different aspects of jet hadronization in the 
presence of a medium.
\\

\begin{figure}[!ht]
\begin{center}
\includegraphics[width=0.55\columnwidth]{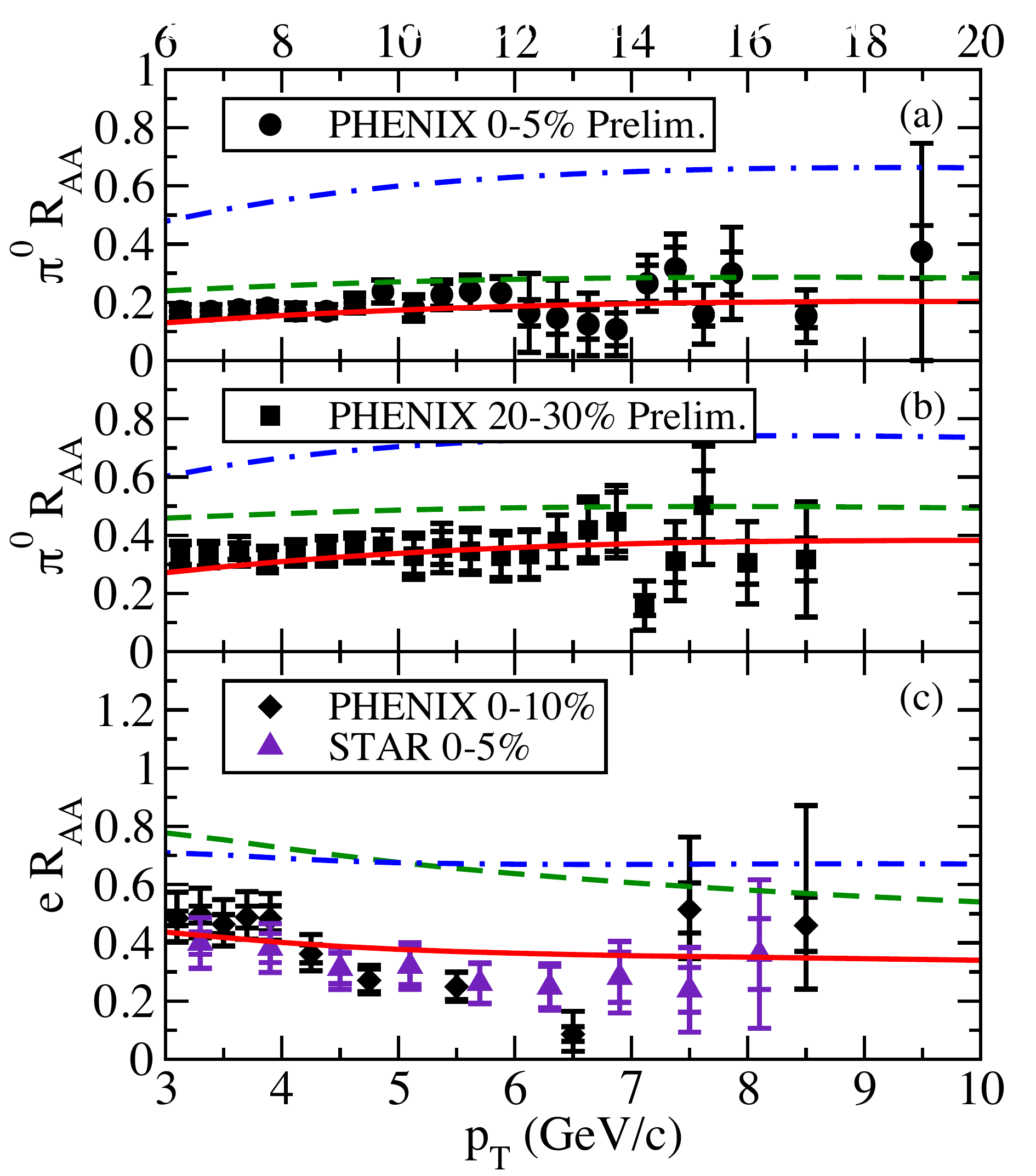}
\caption{A calculation of light and heavy-flavor $R_{AA}$ in an identical medium without assuming weak coupling for the medium degrees of freedom. Data are 
from the PHENIX experiment. Plot reproduced from Ref.~\cite{Qin:2009gw} }
\label{heavy-flavor}
\end{center}
\end{figure}

\emph{ii) Heavy flavor energy loss and SCET in medium:} 
In spite of the successes involved in the theoretical description of light flavor energy loss, the heavy flavor sector has remained a minor 
puzzle. Several attempts at a description of heavy flavor suppression at RHIC energies, using a perturbative description of the plasma were unable to describe the 
$R_{AA}$ of hard semi-leptonic decay electrons that were emitted from the decay of $D$ and $B$ mesons~\cite{Djordjevic:2003zk,Djordjevic:2004nq,Wicks:2005gt}. 
The problem mostly originates from lower energy $b$ quarks (fragmenting to $B$ mesons)~\cite{Renk:2013xaa}.
The disagreement with data 
was considerably improved by the inclusion of a factorized approach and with the eradication of the approximation of a weakly interacting medium~\cite{Qin:2009gw}.
The results of these calculations are shown in Fig.~\ref{heavy-flavor}. 
These calculations are done outside of a fluid dynamical framework and have to be extended to the same level of rigor as the light flavor calculations. 

Beyond this, such 
calculations must also involve the combined effect of the transverse and longitudinal drag on the radiative loss~\cite{Qin:2012fua}. Such calculations for the heavy flavor sector 
are currently underway and involve an entirely new technique, the incorporation of Soft-Collinear-Effective-Theory (SCET)~\cite{Bauer:2000yr,Bauer:2001ct,Bauer:2001yt,Bauer:2002nz} within a medium~\cite{Idilbi:2008vm}. This effective theory, which has had many successes in vacuum applications of pQCD, has recently been adopted by several groups, in an 
attempt to reevaluate and extend the theory of jet modification~\cite{DEramo:2010ak,Fickinger:2013xwa,Ovanesyan:2011kn}. This is a promising new approach that is destined to yield important results in the near future, especially in the domain of 
heavy quark energy loss. 
\\

\emph{iii) Calculation of transport coefficients in lattice QCD:}
In a factorized approach to jet modification, such as in the HT scheme, one factorizes the hard sector of the process involving the hard parton radiated gluons from the 
soft sector, which involves the entities off which the hard parton scatters. This factorization allows one to express the transport coefficients in the medium as an expectation 
of an operator product, as expressed in Eq.~\eqref{qhat_gaugeinvar}. In phenomenological determinations of this transport coefficient, one introduces an additional 
assumption of how $\hat{q}$ (or any other transport coefficient) would scale with the intrinsic properties of the medium, e.g., in Eq.~\eqref{qhat-scaling}. This still requires that the 
overall normalization of the transport coefficient be determined by fitting to one date point. A compendium of such fits from all current approaches may be found in the 
report on $\hat{q}$ from the JET collaboration~\cite{Burke:2013yra}.

However, given the operator definition, it should be possible to determine $\hat{q}$ from first principles, in lattice QCD simulations. Recently, two separate groups have undertaken 
such an effort, from very different directions. In Ref.~\cite{Majumder:2012sh}, the author uses dispersion relations, to relate the physical definition of $\hat{q}$ on the light-cone to 
an ``unphysical'' $\hat{q}$ defined in the deep Euclidean regime. Calculations have so far been carried out in a quenched $SU(2)$  theory. Another approach is that of the 
authors of Ref.~\cite{Panero:2013pla}, where one assumes the applicability of Hard-Thermal-Loop effective theory~\cite{Braaten:1989kk,Braaten:1989mz,Frenkel:1989br,Frenkel:1991ts}, 
and computes the contribution from the hard sector perturbatively and that from the soft sector using a 3-D lattice gauge theory. At this point, it is not possible to compare 
between the two approaches, however there is some disparity between the value of $\hat{q}$ calculated in Ref.~\cite{Panero:2013pla} and that deduced from fits to 
experimental data in Ref.~\cite{Burke:2013yra}. The cause for this is so far unclear and is the subject of current investigations.
\\

\emph{iv)  Renormalization of transport coefficients: }
With the operator expression for $\hat{q}$ factorized from the hard sector, one will encounter the issue of the scale dependence of the factorization. 
Alternatively stated, the operator product that defines $\hat{q}$ will have to be renormalized. It should depend on the scale $\mu_{F}$ at which the 
hard sector is factorized from the soft sector. This is somewhat similar to the DGLAP evolution of parton distribution functions or fragmentation functions. 
For very high energy jets, there should also be an energy (or $x$) dependence in $\hat{q}$. This was first pointed out in Ref.~\cite{CasalderreySolana:2007sw}. 
Recently, there have been a series of calculations to renormalize $\hat{q}$ within the BDMPS approximation~\cite{Liou:2013qya,Blaizot:2014bha,Iancu:2014kga}. There is still no first 
principles calculation for the renormalization of $\hat{q}$ within the HT scheme. A complete calculation of the energy lost by hard partons due to scattering 
induced radiation, will require a first principles calculation of the evolution of transport coefficients with the factorization scale. This will remain true even in the case that 
transport coefficients such as $\hat{q}$ are calculated on the lattice. 
\\

\emph{v) Next-to-Leading-Order calculations:}
With or without the inclusion of non-perturbative evaluations of transport coefficients, the majority of any jet energy loss calculation is perturbative. 
In the parlance of resumed perturbation theory, all jet quenching calculations carried out so far are leading order calculations. To test the convergence of any perturbative 
calculations, one must calculate Next-to-Leading order corrections to these calculations. These are rather involved calculations, and are currently being carried out 
by several groups. The first set of results on NLO corrections to $p_{T}$ broadening, fixed order, have recently appeared~\cite{Kang:2013raa}. 
Closely connected with NLO calculations, is the topic of factorization. So far all calculations tend to calculate the next-to-leading twist correction to LO or NLO process. As such 
there is not conflict with the known violation of factorization at higher twist~\cite{DiLieto:1980dt,Doria:1980ak}. 
\\

\section{Discussions and Conclusions}~\label{apologies}
In this short review, we have attempted to provide a glimpse of the field of jet modification in dense matter for a non-expert. The review is written from a theorist's perspective, 
and as such, is scarce on experimental details. A review from a more experimental perspective, though somewhat older, may be found in Ref.~\cite{dEnterria:2009am}.
We have dealt specifically with perturbative approaches, where the propagation and radiation from the hard parton are treated with pQCD. These are factorized from soft 
matrix elements which encode the properties of the medium which influence the propagation of the jet. 

Starting with Sec.~\ref{1-particle}, we have highlighted the essential features that must be included in any pQCD based description of jet quenching, this was followed by a discussion of the current successes and challenges of MC event generator development. This was followed by a discussion of some of the outstanding problems that are currently being attempted by several groups. In the interest of space, many emerging issues, both experimental and theoretical, e.g., $\gamma$-jet, or $Z$-jet correlations were not addressed in this review. 
The physics of energy loss involved in such observables is rather similar to that discussed above.

\acknowledgments 
The author thanks, R.~Abir and J.~Putschke for helpful discussions, and B.~M\"{u}ller and J.~Schukraft for help with plots. 
This work was supported in part by the U.S. National Science Foundation (grant number PHY-1207918) and by the 
U.S. Department of Energy within the framework of the JET collaboration (grant number DE-SC0004104).


\bibliographystyle{pramana}
\bibliography{refs_pramana_rev}

%
%

\end{document}